\documentclass[
    11pt,
    letterpaper,
    reprint,
    notitlepage,
    superscriptaddress,
    aps,
]{revtex4-2}

\usepackage[dvipsnames]{xcolor}% or package color
\usepackage{mathtools}
\usepackage{graphicx}% Include figure files
\usepackage{dcolumn}% Align table columns on decimal point
\usepackage{bm}% bold math
\usepackage{amsmath,amssymb,amsfonts} 
\usepackage[makeroom]{cancel}
\usepackage{hyperref}
\hypersetup{colorlinks=true,citecolor=blue,linkcolor=blue,urlcolor=blue}
\usepackage{cleveref}
\crefname{equation}{}{}
\crefrangelabelformat{equation}{(#3#1#4--#5#2#6)}
\newcommand{\eref}[1]{Eq.~\eqref{#1}}

\newcommand*{\dotproduct}{\ensuremath{\boldsymbol{\cdot}}}

\newcommand*{\Omq}{\ensuremath{\Omega_{\rm{q}}}}
\newcommand*{\ns}{\ensuremath{n_{\rm{S}}}}
\newcommand*{\nf}{\ensuremath{ n_{\rm{F}}}}
\newcommand*{\gammam}{\ensuremath{\gamma_{\rm{m}}}}
\newcommand*{\omegam}{\ensuremath{\omega_{\rm{m}}}}
\newcommand*{\Omegaf}{\ensuremath{\Omega_{\rm{F}}}}
\newcommand*{\Omegas}{\ensuremath{\Omega_{\rm{S}}}}
\newcommand*{\chiF}{\ensuremath{\chi_{\rm{F}}}}
\newcommand*{\chiS}{\ensuremath{\chi_{\rm{S}}}}

\begin{document}

\title{Characterizing stationary optomechanical entanglement in the presence of non-Markovian noise}

\author{Su Direkci}
    \email{sdirekci@caltech.edu}
    \affiliation{The Division of Physics, Mathematics and Astronomy,
California Institute of Technology, CA 91125, USA} 
\author{Klemens Winkler}
    \affiliation{Vienna Center for Quantum Science and Technology (VCQ), Faculty of Physics, University of Vienna, A-1090 Vienna, Austria}
\author{Corentin Gut}
\thanks{Present address: KEEQuand GmbH, Gebhardtstr. 28, 90762 Fürth, Germany.}
    \affiliation{Vienna Center for Quantum Science and Technology (VCQ), Faculty of Physics, University of Vienna, A-1090 Vienna, Austria}
%\author{Klemens Hammerer}
%    \affiliation{Institute for Theoretical Physics and Institute for Gravitational Physics (Albert-Einstein-Institute),
%Leibniz University Hannover, Appelstrasse 2, 30167 Hannover, Germany}    
\author{Markus Aspelmeyer}
\affiliation{Vienna Center for Quantum Science and Technology (VCQ), Faculty of Physics, University of Vienna, A-1090 Vienna, Austria}
\affiliation{Institute for Quantum Optics and Quantum Information (IQOQI) Vienna, Austrian Academy of Sciences, Boltzmanngasse 3, 1090 Vienna, Austria} 
\author{Yanbei Chen}
\affiliation{The Division of Physics, Mathematics and Astronomy,
California Institute of Technology, CA 91125, USA}

\date{\today}% It is always \today, today,

\begin{abstract}
We study an optomechanical system, where a mechanical oscillator interacts with a Gaussian input optical field. In the linearized picture, we analytically prove that if the input light field is the vacuum state, or is frequency-independently squeezed, the stationary entanglement between the oscillator and the output optical field is independent of the coherent coupling between them, which we refer to as the universality of entanglement. Furthermore, we demonstrate that entanglement cannot be generated by performing arbitrary frequency-dependent squeezing on the input optical field. Our results hold in the presence of general, Gaussian environmental noise sources, including non-Markovian noise.
\end{abstract}

\maketitle

\section{Introduction}

The transition from quantum to classical behavior has been an intriguing question in the past decades. Quantum systems interact with the environment and develop correlations with environmental degrees of freedom, which in turn prohibit observing the quantum nature of the system when viewed by an independent external observer, a phenomenon that is referred to as \textit{decoherence} \cite{Zurek_decoherence,SCHLOSSHAUER2019_decoherence}. Above some rate of decoherence, it is expected that a system starts displaying classical behavior \cite{Quantum-to-classical-Aharonov}.

Optomechanical systems are examples of quantum systems, where a mechanical oscillator interacts with an optical field in the simplest case. In such systems, the momentum exchange between the mechanical and electromagnetic degrees of freedom, a phenomenon that is referred to as radiation pressure \cite{GENES200933, Aspelmeyer_cavity_optomechanics}, gives rise to the entanglement between these two components.
Optomechanical entanglement has applications in quantum sensing \cite{ma_proposal_2017}, quantum control \cite{hofer_entanglement-enhanced_2015, Optomechanical-control-NV,quantum-ground-state}, testing fundamental physics \cite{Probing-macroscopic-quantum, Macroscopic-Quantum-Mechanics-Classical-Spacetime, modulated-optomechanics}, and new quantum technologies \cite{Reversible-quantum-interface, Shandilya2021}.

Optomechanical systems suffer from decoherence due to environmental noise sources, which can destroy or diminish the amount of generated optomechanical entanglement \cite{miao_universal_2010, Effect-of-phase-noise-entanglement, direkci_2024}. 
Examples of such noises are Brownian thermal noise \cite{brownian_thermal_mirrors,Kroker2017}, suspension thermal noise \cite{Hammond2012,Harms2017}, and other technical noises. In this paper, we classify environmental noise sources based on their effect on the mechanical oscillator: we refer to noise sources that get transduced by the mechanical oscillator before being read out by the probe (i.e. the light field) as force noise, and noise sources affecting the probe only as sensing noise, respectively.

In \cite{direkci_2024}, the behavior of optomechanical entanglement was studied assuming a linearized interaction and Gaussian dynamics between an oscillator and the vacuum fluctuations in the input light field \cite{Aspelmeyer_cavity_optomechanics}. Given this system, it has been observed numerically that optomechanical entanglement vanishes above a certain threshold concerning the environmental noise sources only, and cannot be recovered by increasing the interaction strength between the oscillator and the input light field.
In other words, the entangling-disentangling transition is \emph{universal} with respect to the interaction strength. Furthermore, this phenomenon has been observed for environmental noises that are not necessarily Markovian, due to the presence of, for example, structural damping \cite{structural_damping}.

Given this observation, in this paper, we analytically prove the universality of optomechanical entanglement in the presence of general, not necessarily Markovian, environmental noise sources, when the input field is the vacuum state (or is frequency-independently squeezed). Furthermore, we show that if the system is separable when the input field is the vacuum state, entanglement cannot be achieved with arbitrary frequency-dependent squeezing in the input light field.
The universality of the optomechanical entanglement can be interpreted as a quantum-to-classical transition, where the system appears to be ``classical" due to the environment, i.e. it cannot get entangled with the light field since it is strongly correlated with the environment. Calculating the corresponding threshold for environmental noise sources is crucial for upcoming quantum technologies with optomechanical devices, and observing quantum behavior in the macroscopic regime. We also discuss the results in the companion Letter \footnote{See the companion Letter titled ``Universality of Stationary Entanglement in an Optomechanical System Driven by Non-Markovian Noise and Squeezed Light''.}.

This paper is organized as follows. We state the system dynamics, the covariance matrix, and the entanglement criterion in Sec. \ref{sec:system_dynamics}. Then, we demonstrate the universality of the entangling-disentangling transition for arbitrary environmental noise sources in Sec. \ref{sec:universal_entanglement}, assuming that the input light field is in the vacuum state. We relax this assumption and consider arbitrary frequency-dependent squeezing in the input light field in Sec. \ref{sec:fd_squeezing}.

\section{System Dynamics}
\label{sec:system_dynamics}

\begin{figure}
    \centering
\includegraphics[width=0.85\columnwidth]{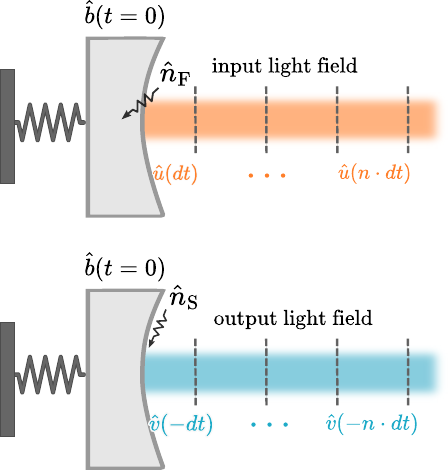}
    \caption{Schematic of the optomechanical system. The mechanical mode $\hat b$ interacts with a light field and reaches a steady state. We refer to the modes of the input light field as $\hat u$, and the reflected (output) light field as $\hat v$, respectively. The schematic shows the equivalence between the spatial and the temporal modes of the light field: for example, the temporal modes of the input field located at the light-mechanics interface for $t>0$ can be thought of as the spatial modes at $t=0$ that have yet to interact with the mirror. Similarly, the temporal modes of the output field for $t<0$ can be thought of as the spatial modes at $t=0$ that have already reflected off the mirror. We label the duration of the temporal modes as $dt$ in the schematic, however, we are interested in the continuum limit, where $dt \rightarrow 0$. Lastly, the mechanical mode is subjected to a force noise $\hat n_{\textrm{F}}$, and the output field is subjected to a sensing noise $\hat n_{\textrm{S}}$.}
    \label{fig:light_curves}
\end{figure}

The system consists of a mechanical oscillator driven by a coherent, input optical field. For a schematic, see Fig. \ref{fig:light_curves}. We work in a suitably displaced frame where the fist moments of the dynamical variables are zero. We denote the position fluctuations of the center of mass of the mechanical oscillator with $\hat{b}_1$, its associated momentum fluctuations with $\hat{b}_2$, the quantum fluctuations in the amplitude (phase) quadratures of the input (output) optical fields with $\hat{u}_1(t)$, $\hat{u}_2(t)$ ($\hat{v}_1(t)$, $\hat{v}_2(t)$), respectively. The dimensionless, linearized equations of motion in Heisenberg picture are \cite{Muller-Ebhardt_2008}
\begin{subequations}
 \label{eqn:eqns_time_domain}
\begin{align}
\hat{v}_1(t) &= \hat{u}_1(t) \,, \\
\hat{v}_2(t) &= \hat{u}_2(t) + \frac{\Omq}{\sqrt{\omegam}} (\hat{b}_1(t)+\hat{n}_{\rm{S}}(t)) \,, \\
 \dot{\hat{b}}_2(t) &= -\gammam \hat{b}_2(t) -\omegam \hat{b}_1(t) +\frac{\Omq}{\sqrt{\omegam}} \hat{u}_1(t) +\hat{n}_{\rm{F}}(t), \label{eqn:b2dot} \\
\dot{\hat{b}}_1(t) &= \omegam \hat{b}_2(t) \,,\label{eqn:b1dot}
\end{align}
\end{subequations}
where $\Omq$ is the characteristic light--mechanics interaction strength, or the coherent coupling \footnote{In terms of system parameters, $\Omq = \sqrt{\frac{I_{\rm{probe}} \omega_{\rm{probe}} }{m c^2}}$, where $I_{\rm{probe}}$ and $\omega_{\rm{probe}}$ are the intensity and carrier frequency of the light probing the mechanics, $c$ is the speed of light in vacuum, and $m$ is the mass of the oscillator.}, $\omegam$ is the mechanical resonance frequency, $\gammam$ is the damping rate of the oscillator, and $\hat{n}_{\rm{F}}(t)$ and $\hat{n}_{\rm{S}}(t)$ are the force and sensing noises, respectively. The commutation relation for the same-time quadrature operators are $[\hat b_1(t), \hat b_2(t)] = 2i $ \footnote{We choose to normalize the quadratures of the mechanical oscillator such that the entries of the commutator matrix, defined in Eq. (\ref{eqn:commutator_matrix}), do not have a factor of 1/2.}.

\begin{table}
\renewcommand{\arraystretch}{1.5}
    \centering
    \begin{tabular}{|c|c|c|}
\hline
    & Force noise & Sensing noise \\
    \hline
   Quantum (measurement) noise & $\hat u_1$  & $\hat u_2$\\
   \hline
   Environmental noise & $ \hat n_{\textrm{F}}$  & $ \hat n_{\textrm{S}}$ \\
   \hline
\end{tabular}
    \caption{\label{tab:noise_classification} Classification of the noise sources affecting the dynamics, given in Eqs. (\ref{eqn:eqns_time_domain}). The noise sources fall under two orthogonal categories, demonstrated by the x and y-axes of the Table. The first category (x-axis) arises due to the nature of the noise: we refer to the fundamental noise that occurs due to the continuous monitoring the oscillator as quantum noise. Conversely, any noise arising due to the environment (e.g. seismic noise, coating thermal noise, etc.) is referred to as environmental noise. The second category (y-axis) describes how the noise source affects the dynamics: a force noise introduces stochastic momentum kicks on the oscillator, whereas a sensing noise is sensed with the position of the oscillator and does not affect its dynamics.}
\end{table}

First, we assume that the driving light (in the displaced frame) is in the vacuum state, with temporally uncorrelated fluctuations. This assumption will be relaxed at the end of the paper, where an arbitrarily frequency-dependent squeezed input light field will be considered. The commutation relations for the quadratures of the input light field are given as $[\hat{u}_{1}(t),\hat{u}_{2}(t')] = 2i \, \delta(t-t')$, $[\hat{u}_{i}(t),\hat{u}_{i}(t')]=0$, $i = 1, 2$. The same relations hold for the output light field. 

As described in the companion Letter \footnotemark[1], the noise sources driving the mechanical oscillator, $\hat u_{1,2},\, \hat n_{\textrm{F}}, \,\hat n_{\textrm{S}}$, can be categorized in two manners, summarized in Table \ref{tab:noise_classification}. First, we refer to the fundamental measurement noise occurring due to the continuous monitoring of the oscillator position as \textit{quantum noise}. This noise is sourced by $\hat u_{1, 2}$ in Eqs. (\ref{eqn:eqns_time_domain}).
Conversely, any additional noise in the system that can in principle be eliminated is referred to as \textit{environmental noise}. This noise is sourced by $\hat n_{\textrm{F}}$ and $\hat n_{\textrm{S}}$ in Eqs. (\ref{eqn:eqns_time_domain}). An orthogonal characterization of the noise sources can be obtained by considering their effect on the system. We refer to the noise sources that produce stochastic momentum kicks to the oscillator as \textit{force noise}, sourced by $\hat u_{1}$ and $\hat n_{\textrm{F}}$. Lastly, the noises that do not affect the dynamics of the oscillator, but are sensed simultaneously with the mechanical position are referred to as \textit{sensing noise}. They are sourced by $\hat u_{2}$, and $\hat n_{\textrm{S}}$.

Examples to $\hat n_{\textrm{F}}$ are, seismic noise \cite{seismic_gw_detectors, correlated_seismic_noise}, suspension thermal noise \cite{suspension_thermal_noise, mirror_suspension_third_gen}, and structural damping \cite{structural_damping, structural_thermal_mirror, torsion_pendulum} in systems with suspended oscillators such as gravitational wave detectors. Further examples include thermal excitations of a substrate in contact with the mechanical oscillator \cite{Electro_Optic_Transducer}, and collision with gas molecules or photon scattering in levitated nanoparticles \cite{gas_photon_recoil, directional_noise_baths, Levitodynamics}. Additionally, examples to $\hat n_{\textrm{S}}$ are, thermal coating noise for coated mirrors \cite{coating_thermal_noise, coating_thermal_gras}, and additional mechanical modes contributing to the outgoing light in membranes or levitated systems \cite{SiN_membrane}. Note that the sensing noise here couples to the detection channel with a rate proportional to the interaction strength $\Omq$ (c.f. Eqs. (\ref{eqn:eqns_time_domain})). There can be other types of sensing noises that do not couple with this rate, e.g. passive or photon losses, and detector dark noise \cite{optomech_entanglement_photon_counting}. We only consider the former type of sensing noise in $\hat n_{\textrm{S}}$, as in the presence of the latter, the universality of optomechanical entanglement does not hold (see Sec. \ref{sec:example}).

Eqs. (\ref{eqn:eqns_time_domain}) describe a system where the mechanical oscillator is driven by the input light field and a force noise, represented by the terms $\frac{\Omq}{\sqrt{\omegam}} \hat{u}_1(t)$ and $\hat{n}_{\rm{F}}(t)$ in Eq. (\ref{eqn:b2dot}), respectively. The system is unconditionally stable and reaches a steady state. Taking the Fourier transform~\footnote{We use the convention $\mathcal{F}\{f(t)\} = \int_{-\infty}^{\infty} f(t)e^{i\omega t} dt$} of Eqs. (\ref{eqn:b2dot}) and (\ref{eqn:b1dot}), the steady state solution for the mechanical oscillator is written as $\hat b_1(\Omega) = \chi (\Omega) F_{ext}(\Omega) $, where $F_{ext}(\Omega) = \Omq \sqrt{\omegam}\hat{u}_1(\Omega) + \omegam\hat{n}_{\rm{F}}(\Omega)$ is the Fourier transform of the sum of external forces driving the oscillator, and $\chi (\Omega) = (-\Omega^2-i\gammam \Omega + \omegam^2)^{-1}$ is the mechanical susceptibility. We refer to the inverse Fourier transform of the mechanical susceptibility as $\chi(t)$, which is a real, causal function of time.
 
\subsection{Structure of the Covariance Matrix}
\label{sec:cov_matrix}

We want to investigate the entanglement between the mechanical oscillator at $t=0$, and the output light field that was reflected from the mirror during $t<0$. For this purpose, we first define the covariance matrix $\mathbf{V}$,
%of the system with $\mathbf{V}_{ij} = \langle \{\hat{X}_i - \langle \hat{X}_i \rangle, \hat{X}_j - \langle \hat{X}_j \rangle \} \rangle/2 $. 
%The Heisenberg uncertainty principle states that $\mathbf{V} + i\mathbf{K}$ is positive semidefinite, where 
which can be written in the following block matrix form
\begin{equation}
\label{eqn:cov_matrix}
\mathbf{V} =  \left[ \begin{array}{cc}
\mathbf{V}^{bb}  & \mathbf{V}^{bv} \\
\mathbf{V}^{vb} & \mathbf{V}^{vv}
\end{array}
\right],\quad \mathbf{V}^{mn} =  \left[ \begin{array}{cc}
V^{mn}_{11}  & V^{mn}_{12} \\
V^{mn}_{21} & V^{mn}_{22} 
\end{array}
\right]
\end{equation}
where $m, n \in \{ v, b \}$, $\mathbf{V}^{bb}$ and $\mathbf{V}^{vv}$ are the covariance matrices of the mechanical mode and the light field respectively, and $\mathbf{V}^{bv}$, $\mathbf{V}^{vb}$ contain the cross-correlations between the mechanical mode and the modes of the optical field, with $\mathbf{V}^{vb} = (\mathbf{V}^{bv})^T$ ($^T$ denotes the transpose). $\mathbf{V}^{bb}$ is a $2 \times 2$ matrix of real numbers, as the mechanical mode is only relevant at the instant of $t=0$. On the other hand, we are interested in the output light field for $t<0$, where the continuous index $t$ labels this infinite set of modes.  $\mathbf{V}^{vv}$ is thus a 2-by-2 matrix of bounded operators on $\mathcal{L}^2 (0, \infty)$, the Hilbert space of square-integrable real functions on the half real line. Similarly, the elements of the 2-by-2 matrix $\mathbf{V}^{vb}$ ($\mathbf{V}^{bv}$) are functions in $\mathcal{L}^2 (0, \infty)$. For example, $V^{bv}_{12}(t)$ is a function denoting the correlations between $\hat b_1(t=0)$ and $\hat v_2(t)$ for $t<0$. To compute the elements of $\mathbf{V}$, we take the inverse Fourier transform of cross spectra between quadratures, as
\begin{align}
\mathbf{V}^{mn}_{kl} &= \frac{1}{2}\langle \{\hat{m}_k(t_1), \hat{n}_l(t_2)\} \rangle \nonumber \\
&= \int_{-\infty}^{+\infty}\frac{d\Omega}{2\pi}S_{m_k n_l}(\Omega) e^{-i\Omega(t_1-t_2)}\,,
\end{align}
where $m, n \in \{ v, b \}$, $k, l \in \{ 1, 2 \}$, $t_{1(2)} = 0$ if $m(n) = b$, $t_{1(2)} < 0$ otherwise, and $S_{m_k n_l}$ is the double-sided cross spectrum between $\hat{m}_k$ and $\hat{n}_l$, given by
\begin{align}
    S_{m_k n_l}(\Omega) \, \delta(\Omega-\Omega') = \frac{1}{4 \pi} \langle \hat m_k(\Omega) \hat n_l^\dagger(\Omega') + \hat n_l^\dagger(\Omega') \hat m_k(\Omega) \rangle
\end{align}
Note that since the input light field is the vacuum, $S_{u_1 u_1} = S_{u_2 u_2} = 1$, and $S_{u_1 u_2} = 0 $. Furthermore, we denote the spectra of $\hat n_{\rm{F}}(t)$ and $\hat n_{\rm{S}}(t)$ with $S_{\nf}(\Omega)$ and $S_{\ns}(\Omega)$, respectively.

First, let us compute $\mathbf{V}^{vv}$. The operators in $\mathbf{V}^{vv}$ are indexed by $\tau \in (-\infty, \infty)$, where $\tau$ increases (decreases) with increasing columns (rows) of the matrix of the operator. $\mathbf{V}^{vv}$ is given as
\begin{align}
\label{eq:vv_cov_vacuum}
\mathbf{V}^{vv} = \begin{bmatrix}
\delta(\tau)  & \Omq^2 \, \chi(-\tau)  \\ \Omq^2 \, \chi(\tau) & \delta(\tau) +\Omq^2\,\chi_{\text{EN}}(\tau) + \Omq^4 \, \Theta(\tau) \end{bmatrix},
\end{align}
where $\Theta(\Omega) = |\chi(\Omega)|^2$ and we define a ``force noise susceptibility" $\chiF(\tau)$, with $\chiF(\Omega) =  \omegam |\chi(\Omega)|^2 S_{\nf}(\Omega)$, and a ``sensing noise susceptibility" $\chiS(\tau)$, with $\chiS(\Omega) = S_{\ns}(\Omega)/ \omegam$ so that , $\chi_{\text{EN}}(\tau) \coloneqq  \chiF(\tau) + \chiS(\tau)$ summarizes the susceptibility of the environmental noises. The matrix representations of $V^{vv}_{ij}, i, j = 1, 2$ are in the Toeplitz form, as their elements depend only on the difference between their respective row and column indices.

Next, let us compute $\mathbf{V}^{bv}$, which will be a matrix of functions indexed by $t$, $t>0$. It contains the correlations between the mechanical oscillator and the light field:
\begin{align}
\label{eqn:cov_light_mass}
\mathbf{V}^{bv} = \sqrt{\omegam} \Omq \begin{bmatrix}
\chi(t)  & \chiF(t) + \Omq^2 \, \Theta(t) \\ \frac{1}{\omegam} \dot \chi(t) & \frac{1}{\omegam} \dot \chi_{\rm{F}}(t) + \frac{\Omq^2}{\omegam} \dot \Theta(t)  \end{bmatrix}  \,,
\end{align}   
where $t$ increases with increasing columns. Since $\hat b_2(t) = \dot{\hat{b}}_1(t)/\omegam$ from Eq.~\eqref{eqn:b1dot}, the cross-correlations involving $\hat b_2(t)$ are derivatives of the cross-correlations involving $\hat b_1(t)$, normalized by $\omegam$. The terms containing $\Theta$ in the second columns correspond to the amplitude fluctuations of the input light, transduced twice by the mechanical oscillator (recall that $\Theta$ is $|\chi|^2$ in the frequency domain). %We observe that $V^{bv}_{i2}$, $i= 1,2$, contains a term due to the amplitude fluctuations in the input light field driving the oscillator. 
These fluctuations 
%are transduced by the oscillator, and 
emerge as noise in the phase quadrature of the output light field. This phenomenon 
is referred to as \emph{back action}.
%in the literature.

Lastly, let us compute $\mathbf{V}^{bb}$, which is a matrix of real numbers:
\begin{subequations}
\label{eqn:mech_cov_matrix}
\begin{align}
&\mathbf{V}^{bb} = \left[ \begin{array}{cc}
V^{bb}_{11} & 0 \\
0 & V^{bb}_{22} \end{array} \right]  \\
& V^{bb}_{11} = \omegam \left( \chiF(t=0) + \Omq^2\, \Theta(t=0) \right)  \\
& V^{bb}_{22} = \frac{-1}{\omegam} \left(  \ddot \chi_{\rm{F}}(t=0) + \Omq^2\, \ddot \Theta(t=0) \right)
\end{align}
\end{subequations}
This matrix is diagonal due to vanishing cross-correlations between $\hat b_1(t)$ and $\hat b_2(t)$ at equal times.

\subsection{Entanglement Criterion}
\label{sec:entanglement_criterion}

We are interested in the bipartite entanglement between the mechanical oscillator and the continuum of temporal modes of the output light. 
Since we assume Gaussian dynamics and noise sources, the state of the mechanical mode and the temporal modes of the light are Gaussian. In this configuration, the PPT criterion is necessary and sufficient for verifying the presence of bipartite entanglement \cite{peres_separability_1996,simon_peres-horodecki_2000, duan_inseparability_2000, werner_bound_2001}. For continuous variables, the criterion asserts that a state is separable iff $\mathbf{V}_{\text{pt}} + i\mathbf{K} \geq 0 $; $\mathbf{V}_{\text{pt}}$ denotes the partial transpose of $\mathbf{V}$ with respect to one of the parties \cite{simon_peres-horodecki_2000, adesso_entanglement_2007}, and 
$\mathbf{K}$ is the commutator matrix, defined as
\begin{equation}
\mathbf{K} = \left[
\begin{array}{ccc}
\mathbf{K}^b \\
& \mathbf{K}^v
\end{array}
\right] \,,
\end{equation}
with
\begin{align}
\label{eqn:commutator_matrix}
\mathbf{K}^b =
\left[\begin{array}{cc}
    0 & 1 \\
    -1 & 0
    \end{array}\right],\;
    \mathbf{K}^v =
     \left[\begin{array}{cc}
    0 & \delta(t) \\
    -\delta(t) & 0
    \end{array}\right] \,.
\end{align}

In our configuration, 
%we prefer to partially transpose the mechanical part, which 
the partial transpose operation amounts to reversing the sign of the momentum of the mechanical mode: \mbox{$\hat b_2(t) \rightarrow - \hat b_2(t)$}, which also puts a minus sign on the rows and columns of $\mathbf{V}$ associated with $\hat b_2(t)$. For example,
\begin{align}
\label{eqn:v_pt_general}
\mathbf{V}^{bv}_{\text{pt}} = \begin{bmatrix}
        \;\;\; V^{bv}_{11}(t) & \;\;\;V^{bv}_{12}(t) \\[1mm]
        -V^{bv}_{21}(t) & -V^{bv}_{22}(t)
    \end{bmatrix}.
\end{align}
%Additionally, the degree of violation of the positive definiteness above provides a quantitative measure of entanglement given by the positive values of the \emph{logarithmic negativity} \cite{vidal_computable_2002}
%\begin{equation}
%\label{eqn:negativity}
%E_\mathcal{N} = \sum_{j} \text{max}\{0, -\text{log}_2 (\lambda_{j})\}.
%\end{equation}
%The $\lambda_{j}$ %are the positive eigenvalues of $i \mathbf{K} \mathbf{V}_{\text{pt}}$ and 
%are the \emph{symplectic eigenvalues} of $\mathbf{V}$, which are the positive solutions of the following generalized eigenvalue problem,
%\begin{align}
%\label{eqn:symp_eigenvalue}
%    \mathbf{V}_{\rm pt} \mathbf{v} &= i \lambda %\mathbf{K} \mathbf{v}.
%\end{align}
Since we are looking for entanglement of one of the modes against the rest, $\mathbf{V}_{\text{pt}} + i\mathbf{K}$ can have at most one negative eigenvalue \cite{serafini_quantum_2017, adesso_entanglement_2007}. Therefore, the optomechanical entanglement exists if the following condition holds:
\begin{align}
\label{eqn:determinant_full}
\mathrm{det} \left[ \mathbf{V}^{bb} + i \mathbf{K}^b - \mathbf{V}^{bv}_{\text{\text{pt}}} \boldsymbol{\cdot} (\mathbf{V}^{vv} + i \mathbf{K}^v)^{-1} \boldsymbol{\cdot} \mathbf{V}^{vb}_{\text{\text{pt}}} \right] < 0. 
\end{align}
Note that the dot product $\boldsymbol{\cdot}$ is defined as, for an $n$-by-$n$ matrix of operators $\mathbf{A}$, an $n$-by-$m$ ($m$-by-$n$) matrix of functions $\mathbf{f}$ ($\mathbf{g}$), 
\begin{align}\label{eqn:dot_prod}
    &(\mathbf{g} \boldsymbol{\cdot} \mathbf{A} \boldsymbol{\cdot} \mathbf{f})_{kl} \nonumber \\ &= \sum_{i,j} \int_0^{\infty} \int_0^{\infty} dt\,dt'\,g_{ki}(t') A_{ij}(t-t') f_{jl}(t).
\end{align}
From Eq. (\ref{eqn:determinant_full}), the existence of optomechanical entanglement depends on a complex trade-off between the strength of the force noise, the sensing noise, and the quantum noise due to vacuum fluctuations. We explore this relation in depth in the following sections. 

Lastly, entanglement is quantified with the \emph{logarithmic negativity}, defined as \cite{negativity, plenio_negativity}
\begin{align}
    E_\mathcal{N} = \sum_j \text{max}\{0, -\text{log}(\lambda_j)\}
\end{align}
where the sum is over the symplectic eigenvalues $\lambda_j$ of the partially transposed covariance matrix, and a non-zero logarithmic negativity indicates the presence of optomechanical entanglement.

\section{Universal Entanglement}
\label{sec:universal_entanglement}

We first simplify the condition to observe entanglement in Eq. (\ref{eqn:determinant_full}) in Sec. \ref{sec:entanglement_indicator}. Subsequently, we demonstrate the universality of the entangling-disentangling transition in Sec. \ref{sec:proving_universality}. Lastly, we give an example to how this transition constrains the amount of environmental noise allowable in order to observe optomechanical entanglement in Sec. \ref{sec:example}.

\subsection{Entanglement Indicator}
\label{sec:entanglement_indicator}

To simplify Eq. (\ref{eqn:determinant_full}), we first evaluate $\left(\mathbf{V}^{vv}_{\text{pt}} + i \mathbf{K}^v \right)^{-1}$. Using standard properties of the 2-by-2 block-matrices (see Appendix \ref{app:schur}), we write
\begin{subequations}
\label{eqn:inverse_schur}
\begin{align}
&(\mathbf{V}^{vv}_{\text{pt}} + i \mathbf{K}^v )^{-1} = \mathbf{\Sigma} \dotproduct \mathbf{D}_M \dotproduct \mathbf{\Sigma}^\dagger + \mathbf{\Delta}, \label{eqn:breaking_down_big_matrix_general} \\
 & \mathbf{\Delta} = \left[ \begin{array}{cc}
 V^{vv}_{11}(\tau)^{-1}  & 0 \\
 0 & 0  \end{array} \right], \\
&\mathbf{\Sigma} = \begin{bmatrix}
     V^{vv}_{11}(\tau)^{-1}  & 0 \\
 0 & \delta(\tau)
\end{bmatrix} \dotproduct \begin{bmatrix}
    V^{vv}_{12}(\tau) + i \delta(\tau) & 0 \\ -\delta(\tau) & 0 
\end{bmatrix}, \label{eqn:V_M_defn_1} \\
& \mathbf{D}_M = \begin{bmatrix}
    M^{-1}(\tau) & 0 \\
    0 & M^{-1}(\tau)
\end{bmatrix},
\end{align}
\end{subequations}
with \footnote{Note that in general, the operator $M$ is a function of two time indices, in the form of $M(t, t')$. Here, assume $M(\tau) = M(t-t')$ for simplicity, which is true when the input light field is the vacuum state.}
%\begin{align}
%\label{eq:m_defn_general}
%    M(t, t') &= V^{vv}_{22}(t, t') - \left[i \lambda + V^{vv}_{21}(t, t')\right] \dotproduct \nonumber \\ &V^{vv}_{11}(t, t')^{-1} \dotproduct \left[ -i \lambda + V^{vv}_{12}(t, t') \right]
%\end{align}
\begin{align}
\label{eq:m_defn_general}
    M(\tau) = & V^{vv}_{22}(\tau) - \left[-i \delta(\tau) + V^{vv}_{21}(\tau)\right] \dotproduct V^{vv}_{11}(\tau)^{-1} \nonumber \\ & \dotproduct \left[ i \delta(\tau) + V^{vv}_{12}(\tau) \right]
\end{align}
and $M^{-1}(\tau)$ is an operator calculated with the Wiener-Hopf method; see Appendix \ref{app:wiener-hopf} for a summary. 
Plugging in $V^{vv}_{ij}(t)$, $i, j = 1, 2$ from Eq. (\ref{eq:vv_cov_vacuum}), and using the dot product convention in Eq. (\ref{eqn:dot_prod}), we obtain
\begin{subequations}
\begin{align}
( \mathbf{V}^{vv}_{\text{pt}} & + i \mathbf{K}^v )^{-1} = \mathbf{\Sigma} \dotproduct \mathbf{D}_M \dotproduct \mathbf{\Sigma}^\dagger + \left[ \begin{array}{cc}
\delta(\tau)  & 0 \\
 0 & 0  \end{array} \right], \label{eqn:breaking_down_big_matrix} \\
    \mathbf{\Sigma} = & \begin{bmatrix}
    \Omq^2 \, \chi(-\tau)+i \delta(\tau) & 0 \\[1mm] -\delta(\tau) & 0 
\end{bmatrix}, \\
M(\tau) = & -i \Omq^2 \left(\chi(\tau) - \chi(-\tau) \right) + \Omq^2 \chi_{\text{EN}}(\tau) \label{eqn:m_vac_defn}
\end{align}
\end{subequations}
When the second operator in Eq. (\ref{eqn:breaking_down_big_matrix}) operates on $\mathbf{V}^{bv}_{\text{pt}}$ and $\mathbf{V}^{vb}_{\text{pt}}$ in \eref{eqn:determinant_full}, we obtain
\begin{subequations}
\begin{align}
& \int_0^{\infty} dt \left[ \begin{array}{cc}
V^{bv}_{11}(t)^2  & -V^{bv}_{11}(t) V^{bv}_{21}(t) \\[1mm]
-V^{bv}_{21}(t)V^{bv}_{11}(t) & V^{bv}_{21}(t)^2  \end{array} \right] \\ &= \int_{0}^{\infty} dt \, \Omq^2 \begin{bmatrix}
    \omegam  \chi(t)^2 & - \chi(t) \dot \chi(t) \\[2pt]
    - \chi(t) \dot \chi(t)  & \frac{1}{\omegam} \dot \chi(t)^2
\end{bmatrix} \label{subeq:18b} \\ &= \int_{-\infty}^{\infty} dt \, \Omq^2 \begin{bmatrix}
    \omegam  \chi(t)^2 & - \chi(t) \dot \chi(t) \\[2pt]
    - \chi(t) \dot \chi(t)  & \frac{1}{\omegam} \dot \chi(t)^2
\end{bmatrix} \label{subeq:18c} \\ &= \int_{-\infty}^{\infty} d\Omega \, \Omq^2 \begin{bmatrix}
    \omegam  |\chi(\Omega)|^2 & 0 \\[2pt]
    0  & \frac{\Omega^2}{\omegam} |\chi(\Omega)|^2
\end{bmatrix} \label{subeq:18d} \\ &= \Omq^2 \begin{bmatrix}
    \omegam \Theta(0) & 0 \\[2pt]
    0  & \frac{-1}{\omegam} \ddot \Theta(0)
\end{bmatrix} \label{subeq:18e},
\end{align}
\end{subequations}
since $\int_{0}^{\infty} dt \, \chi(t) \dot \chi(t) = -\chi(0)^2/2 = 0$. We were able to extend the lower integration limit to $-\infty$ in Eq. (\ref{subeq:18c}) due to the causality of $\chi(t)$. Then, we used the Plancherel theorem to go from the time domain to the frequency domain to calculate the diagonal elements in Eq. (\ref{subeq:18d}). Lastly, we expressed the diagonal elements as inverse Fourier transforms at time $t=0$ in Eq. (\ref{subeq:18e}). We notice that this expression is contained in $\mathbf{V}^{bb}$ [Eq. (\ref{eqn:mech_cov_matrix})], therefore it cancels out of the determinant in Eq. (\ref{eqn:determinant_full}), which becomes
\begin{align}
\label{eq:determinant_after_vacuum_cancellation}
&\mathrm{det} \left[ \widetilde{\mathbf{V}}^{bb} + i \mathbf{K}^b - \mathbf{V}^{bv}_{\text{pt}} \dotproduct  \mathbf{\Sigma} \dotproduct \mathbf{D}_M \dotproduct \mathbf{\Sigma}^\dagger  \dotproduct ({\mathbf{V}^{bv}_{\text{pt}}})^T \right] < 0, \nonumber \\ &\widetilde{\mathbf{V}}^{bb} \coloneqq \left[ \begin{array}{cc}
\omegam \chiF(0) & 0 \\
0 & \frac{-1}{\omegam} \ddot \chi_{\rm{F}}(0) \end{array} \right]
\end{align}
where we rewrote $\mathbf{V}^{vb}_{\text{pt}}$ as $({\mathbf{V}^{bv}_
{\text{pt}}})^T$. We refer to this simplification as the \textit{vacuum cancellation}, since the variances of the quadratures of the mechanical oscillator due to the vacuum fluctuations factor out of the inequality in Eq. (\ref{eqn:determinant_full}).

To calculate $\mathbf{V}^{bv}_{\text{pt}} \dotproduct\mathbf{\Sigma} \dotproduct \mathbf{D}_M \dotproduct \mathbf{\Sigma}^\dagger  \dotproduct ({\mathbf{V}^{bv}_{\text{pt}}})^T$, we first operate with $\mathbf{\Sigma}$ on $\mathbf{V}^{bv}_{\text{pt}}$ from the right,
\begin{align}
\label{eqn:inner_prod_v_m_v_pt_full}
&\begin{bmatrix}
        \;\;\, V^{bv}_{11}(t) & \;\;\, V^{bv}_{12}(t) \\[1mm]
        -V^{bv}_{21}(t) & -V^{bv}_{22}(t)
    \end{bmatrix} \dotproduct \begin{bmatrix}
    \Omq^2 \, \chi(-\tau)+i\delta(\tau) & 0 \\ -\delta(\tau) & 0 
\end{bmatrix}.
\end{align}
Here, we use the index $t > 0$ to indicate functions, and $\tau \in (-\infty, \infty)$ to indicate operators. We use this convention throughout the rest of the paper. Performing the matrix product, we obtain
\begin{align}
\label{eqn:inner_prod_v_m_v_pt_vac}
    \begin{bmatrix}
      \;\;\, i V^{bv}_{11}(t) + \Omq^2 \, V^{bv}_{11}(t) \dotproduct \chi(-\tau) - V^{bv}_{12}(t) & 0 \\[1mm]  -i V^{bv}_{21}(t) - \Omq^2 \, V^{bv}_{21}(t) \dotproduct \chi(-\tau) + V^{bv}_{22}(t) & 0
    \end{bmatrix}
\end{align}
where we used the fact that taking the dot product with the identity operator $\delta(\tau)$ does not modify the functions.
Taking the dot product between $V^{bv}_{11}(t)$ and $\chi(-\tau)$ and going to the frequency domain, we have
\begin{align}
    \left( V^{bv}_{11}(t) \dotproduct \chi(-\tau) \right)(\Omega) &= \mathcal{F} \biggl\{ \int_0^{\infty} dt' \chi(t'-t) V^{vb}_{11}(t') \biggr\} 
    \nonumber \\ &= \Omq \sqrt{\omegam} \mathcal{F} \biggl\{ \int_{-\infty}^{\infty} dt' \chi(t'-t) \chi(t') \biggr\} 
    \nonumber \\ &= \Omq \sqrt{\omegam} \chi(\Omega)^\dagger \chi(\Omega)\nonumber \\ &= \Omq \sqrt{\omegam} |\chi(\Omega)|^2
\end{align}
where we used the definition of $V^{bv}_{11}$ given in Eq. (\ref{eqn:cov_light_mass}), and we extended the integration boundary using that $\chi(t) = 0$ for $t<0$. Then, we find that 
\begin{align}
    V^{bv}_{11}(t) \dotproduct \chi(-\tau) &= \sqrt{\omegam} \Omq \mathcal{F}^{-1}\bigl\{ |\chi(\Omega)|^2 \bigr\} \nonumber \\
    &= \sqrt{\omegam}\Omq \Theta(t)
\end{align}
Similarly, we can also write $V^{bv}_{21}(t) \dotproduct \chi(-\tau) = \sqrt{\omegam} \Omq \dot \Theta(t)$. Substituting these expressions back in Eq. (\ref{eqn:inner_prod_v_m_v_pt_vac}), we obtain
\begin{align}
\begin{bmatrix}
      \;\;\,i V^{bv}_{11}  - \widetilde{V}^{bv}_{12} & 0 \\ -i V^{vb}_{21} + \widetilde{V}^{vb}_{22} & 0
    \end{bmatrix}
\end{align}
where we have defined
\begin{align}
\label{eqn:force_noise_vb_redefined}
    \widetilde{V}^{bv}_{12}(t) =  \sqrt{\omegam} \Omq \, \chiF(t), \;\; \widetilde{V}^{bv}_{22}(t) = \frac{\Omq}{\sqrt{\omegam}} \dot \chi_{\rm{F}}(t)
\end{align}
i.e. we redefine $\widetilde{V}^{bv}_{i2}(t)$, $i=1,2$ such that they only contain the correlations due to the force noise, and not the back action.

Similar calculations allow to re-express $\mathbf{V}^{bv}_{\text{pt}} \dotproduct  \mathbf{\Sigma} \dotproduct \mathbf{D}_M \dotproduct \mathbf{\Sigma}^\dagger  \dotproduct ({\mathbf{V}^{bv}_{\text{pt}}})^T$ as
\begin{equation}
    \widetilde{\mathbf{V}}^{bv}_{\text{pt}}(t') \dotproduct \mathbf{V}_M(\tau=t-t') \dotproduct (\widetilde{\mathbf{V}}^{bv}_{\text{pt}}(t))^T
\end{equation}
with
\begin{align}
\label{eqn:v_m_definition}
\widetilde{\mathbf{V}}^{bv}_{\text{pt}} \coloneqq \begin{bmatrix}
    \;\;\, V^{bv}_{11} & \;\;\, \widetilde{V}^{bv}_{12} \\[1mm] -V^{bv}_{21} & -\widetilde{V}^{bv}_{22}
\end{bmatrix}, \mathbf{V}_M \coloneqq \begin{bmatrix}
     M^{-1} & -iM^{-1} \\[1mm] iM^{-1} & \quad\; M^{-1}
\end{bmatrix}.
\end{align}
Therefore, we can rewrite Eq. (\ref{eq:determinant_after_vacuum_cancellation}) and define the following \textit{entanglement indicator}
\begin{align}
\label{eqn:entanglement_indicator}
    \mathrm{det} & \left[ \widetilde{\mathbf{V}}^{bb} + i\mathbf{K}^b - \widetilde{\mathbf{V}}^{bv}_{\text{pt}} \dotproduct \mathbf{V}_M \dotproduct (\widetilde{\mathbf{V}}^{bv}_{\text{pt}})^T  \right] < 0  \nonumber \\ &\iff \text{optomechanical entanglement}.
\end{align}
This inequality is our starting point for the subsequent considerations.

\subsection{Demonstrating Universality
}
\label{sec:proving_universality}

Given the entanglement indicator in Eq. (\ref{eqn:entanglement_indicator}), we explore the existence of entanglement for different configurations of environmental noise sources. For example, we first assume a very large sensing noise,
i.e. the limit of $S_{\ns} \rightarrow \infty$. In this limit, $\chiS(t) \rightarrow \infty$, therefore $M \rightarrow \infty$, and consequently $M^{-1} \rightarrow 0$. However, the elements of $\mathbf{V}^{bv}_{\text{pt}}$ are finite, since they do not contain the correlations due to the sensing noise. Then, the rightmost term in Eq. (\ref{eqn:entanglement_indicator}) goes to zero, and the entanglement indicator is modified to
\begin{align}
\label{eqn:no_sensing_noise}
    &\mathrm{det} \left[ \widetilde{\mathbf{V}}^{bb} + i\mathbf{K}^b \right] < 0  \iff \text{entanglement.}
\end{align}
$\widetilde{\mathbf{V}}^{bb}$ contains the fluctuations in the quadratures of the mechanical oscillator due to the force noise only. Therefore, this matrix can be thought as the covariance matrix of an oscillator in the absence of any optomechanical interaction, subjected to force noise. Since this is a physical covariance matrix, $\widetilde{\mathbf{V}}^{bb} + i\mathbf{K}^b \geq 0$ from the Heisenberg uncertainty principle. Therefore, Eq. (\ref{eqn:no_sensing_noise}) cannot be realized for any physical force noise spectrum, and there can be no optomechanical entanglement in this limit.

Next, we investigate entanglement in the absence of sensing noise. The condition in Eq. (\ref{eqn:entanglement_indicator}) reduces to the following expression for a high-Q oscillator ($\omegam \gg \gammam$) 
\begin{align}
    S_{\nf}(\omegam) \geq 2 \gammam \iff \text{entanglement.}
\end{align}
We derive this inequality in Appendix \ref{app:no_sensing_noise}, which holds for all force noise spectra by the fluctuation-dissipation theorem \cite{callen_1951, Kubo_1966}. Therefore, the mechanical oscillator is entangled with the output light field for any force noise spectrum in the absence of sensing noise.

Summarizing the two scenarios discussed above, for a given force noise spectrum, the existence of entanglement depends on the amount of sensing noise present in the system, signifying that there exists an entangling-disentangling transition at which optomechanical entanglement emerges. Furthermore, this transition is unique for a given force noise and sensing noise spectrum, as demonstrated in Appendix \ref{app:uniqueness}.

Furthermore, from Eqs. (\ref{eqn:m_vac_defn}), (\ref{eqn:force_noise_vb_redefined}), (\ref{eqn:v_m_definition}), and (\ref{eqn:entanglement_indicator}), \mbox{$M(t) \propto \Omq^{2}$}, and $\widetilde{\mathbf{V}}^{bv}_{\text{pt}} \propto \Omq$. Therefore, the term $\widetilde{\mathbf{V}}^{bv}_{\text{pt}}  \dotproduct \mathbf{V}_M \dotproduct (\widetilde{\mathbf{V}}^{bv}_{\text{pt}})^T$ in Eq. (\ref{eqn:entanglement_indicator}) is independent of $\Omq$. We also have that $\widetilde{\mathbf{V}}^{bb}$ and $\mathbf{K}^b$ are already independent of $\Omq$. Therefore, $\Omq$ factors out of Eq. (\ref{eqn:entanglement_indicator}), making the entangling-disentangling transition independent of the interaction strength between the mechanical oscillator, and the output light field. Therefore, this transition is universal with respect to the interaction strength. To restate our definition of universality, we call a transition universal if it i) exists for a finite level of sensing noise, ii) is unique with respect to the amount of sensing noise, and iii) is independent of the coherent coupling $\Omq$.

After the transition (i.e. for an entangled state), we saw numerically that the negativity increases with $\Omq$, which supports our intuition that the optomechanical entanglement is strengthened with an increased interaction strength. However, most interestingly, the universality implies that, for a noise configuration before the transition (i.e. for a separable state), one cannot make the state entangled by increasing the interaction strength.

In Appendix \ref{app:yanbei_proof}, we demonstrate another way to reach Eq. (\ref{eqn:entanglement_indicator}), which provides more physical intuition. That derivation suggests that the entanglement condition in Eq. (\ref{eqn:entanglement_indicator}) reduces to testing whether the conditional quantum state of a mechanical oscillator driven solely by a force noise $\hat{n}_{\rm{F}}$ satisfies the Heisenberg uncertainty principle. The state is conditioned on the output light field, which senses the oscillator with a sensing noise $\hat{n}_{\rm{S}}$. If the uncertainty principle is violated, optomechanical entanglement will be observed. We notice that the mechanical oscillator in this picture in the alternative derivation is not driven by vacuum fluctuations, which is due to the vacuum cancellation shown in Sec. \ref{sec:proving_universality}.

\subsection{Example Case}
\label{sec:example}

To derive a simple analytical bound on the maximum allowable environmental noise in order to observe optomechanical entanglement, we assume white noise spectra parametrized with $\Omegaf$ and $\Omegas$ for the force and the sensing noise, respectively; cf \cite{chen_macroscopic_2013, direkci_2024}. More specifically, we assume that their spectra are in the form of 
\begin{align}
\label{eq:white_force_sensing}
    S_{\nf}(\Omega) = \frac{2\Omegaf^2}{\omegam}, \quad S_{\ns}(\Omega) = \frac{2\omegam}{\Omegas^2}.
\end{align}
Consequently, \mbox{$\chiF(\Omega) = 2\Omegaf^2 |\chi(\Omega)|^2$}, and \mbox{$\chiS(\Omega) = 2/\Omegas^2$}.

Furthermore, we assume the \textit{free-mass limit}, i.e. the mechanical resonance frequency $\omegam$ of the oscillator is much smaller than the other frequencies in the system. In this limit, $\Omegaf, \Omegas \gg \omegam, \gammam$. Lastly, we assume the presence of passive losses occurring due to e.g. a photodetector with non-unity efficiency, whose effect on the dynamics is described in Appendix \ref{app:passive_loss}. Even though these losses act effectively as a sensing noise on the system, they do not couple to the detection channel with a rate of $\Omq$, different from the sensing noise $\hat n_{\textrm{S}}$ in Eqs. (\ref{eqn:eqns_time_domain}). Hence, in the presence of such losses, optomechanical entanglement is no longer universal with respect to the coherent coupling $\Omq$.

\begin{figure}
    \centering
\includegraphics[width=\linewidth]{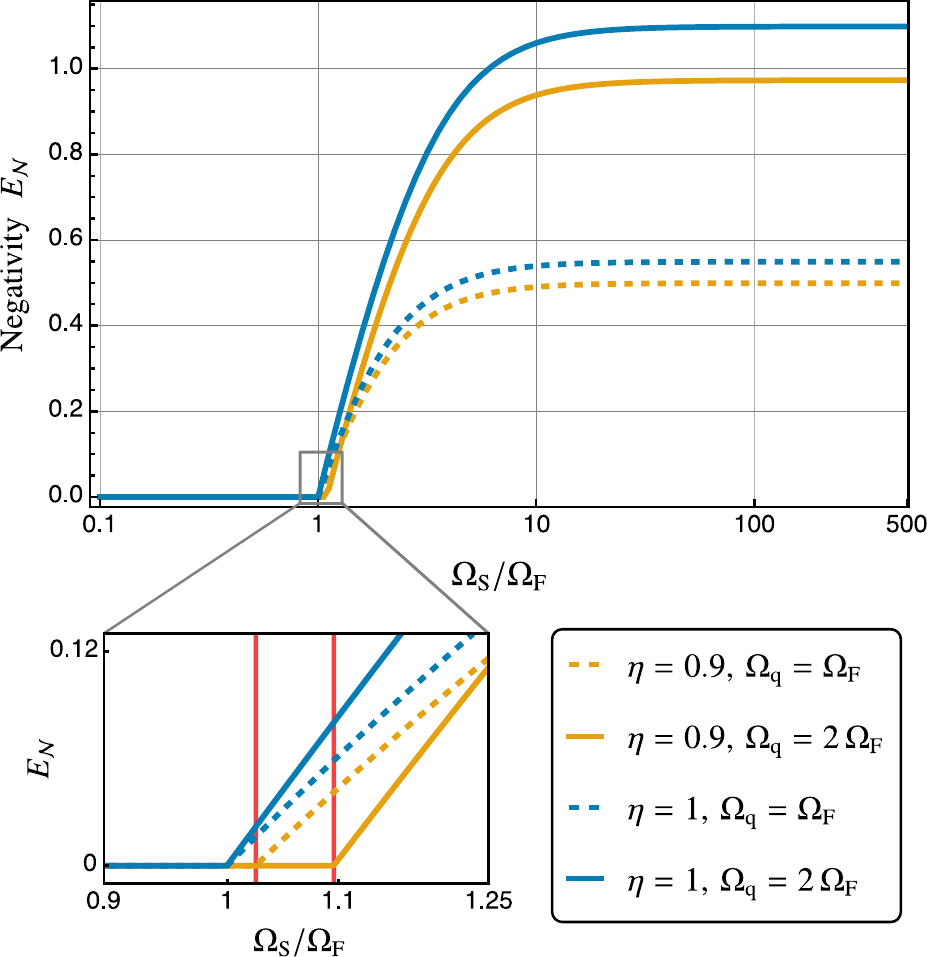}
\caption{Logarithmic negativity for the case where the input light field is the vacuum state, for a photodetector efficiency of $\eta = 0.9$ (plotted in yellow), and $\eta = 1$ (plotted in blue), and white environmental noise sources. We fix the force noise level with $\Omegaf/(2\pi) = 10$ Hz, and vary the sensing noise level in the range of $\Omegas/(2\pi) \in [1,\, 500]$ Hz. The parameters of the mechanical oscillator are chosen as $\omegam/(2\pi) = 1$ Hz and $\gammam/(2\pi) = 0.01$ Hz. We observe that for a unity photodetector efficiency ($\eta = 1$), the entangling transition is characterized by $\Omegaf \approx \Omegas$, as observed numerically in Ref. \cite{direkci_2024}. For a non-unity photodetector efficiency ($\eta <1$), the entangling transition is not universal with respect to the interaction strength $\Omq$. For this case, the transition is characterized by $\sqrt{\Omegaf^2 + (1-\eta) \Omq^2/2} \approx \Omegas$, drawn by the red vertical lines in the inset.}
    \label{fig:dark_noise}
\end{figure}

For a white force and sensing noise, and a photodetector efficiency of $\eta < 1$, the spectrum of $M$ is given by (Eq. (\ref{eqn:m_vac_defn}))
\begin{align}
    \frac{M(\Omega)}{2\eta\,\Omq^2} = \frac{1}{\Omegas^2} +  |\chi(\Omega)|^2 \left[\Omegaf^2 + \gammam \Omega + \frac{\Omq^2}{2}(1-\eta) \right].
\end{align}
In order to apply the Wiener-Hopf method, we need to write $M(\Omega)$ as a product of its causal and anti-causal poles and zeros, i.e. $M(\Omega) = M(\Omega)_+M(\Omega)_-$, where $M(\Omega)_+ = M(\Omega)_-^\dagger$, and $M(\Omega)_+$ ($M(\Omega)_-$) contains the causal (anti-causal) poles and zeros. In the free-mass limit, $M(\Omega)_+$ is given up to first order in $\omegam, \gammam$ as
\begin{align}
    M(\Omega)_+ \approx \frac{\sqrt{2\eta} \, \Omq}{\Omegas} \cdot \frac{(\Omega-\Omega_{z,+})(\Omega-\Omega_{z,-})}{\Omega^2+i\gammam\Omega - \omegam^2},
\end{align}
where
\begin{align}
    \Omega_{z, \pm} &= \frac{-i\pm 1}{\sqrt{2}} \sqrt{\Omegas \Omega_{\textrm{N}}} \mp \frac{i \Omegas \gammam}{4\Omega_{\textrm{N}}}.   
\end{align}
with $\Omega_{\textrm{N}} = \sqrt{\Omegaf^2 + (1-\eta) \Omq^2/2}$.
To evaluate the entanglement indicator in Eq. (\ref{eqn:entanglement_indicator}), we need to obtain the expressions for $\widetilde{\mathbf{V}}^{bb}$ and $\widetilde{\mathbf{V}}^{bv}_{\text{pt}}$. From Eqs. (\ref{eq:determinant_after_vacuum_cancellation}) and  (\ref{eqn:force_noise_vb_redefined}), we calculate
\begin{align}
    \widetilde{\mathbf{V}}^{bb} &= \frac{\Omega_{\textrm{N}}^2}{\gammam \omegam} \begin{bmatrix}
        1 & 0 \\
        0 & 1
    \end{bmatrix}, \nonumber \\
\widetilde{\mathbf{V}}^{bv}_{\text{pt}}(t) &= \sqrt{\omegam \eta} \, \Omq \begin{bmatrix}
\chi(t)  & 2\, \Omega_{\textrm{N}}^2 \,\Theta(t) \\[1.5mm] \frac{-1}{\omegam} \dot \chi(t) & \frac{-1}{\omegam} 2\, \Omega_{\textrm{N}}^2 \, \dot  \Theta(t)  \end{bmatrix}
\end{align}
where $\Theta(\Omega) = |\chi(\Omega)|^2$. Proceeding with the Wiener-Hopf method, we obtain up to the terms that contribute to constant  order in $\omegam, \gammam$
%\begin{widetext}
%{\textcolor{blue}{
%\begin{align}
%\widetilde{\mathbf{V}}^{bv}_{\text{pt}} \dotproduct  \mathbf{V}_M \dotproduct
%(\widetilde{\mathbf{V}}_\text{pt}^{bv})^T \approx  \begin{bmatrix}
%        \frac{\Omega_{\textrm{N}}^2}{\gammam \omegam} + 2^{3/2} \omegam \sqrt{\frac{\Omega_{\textrm{N}}}{\Omegas^3}} & -i \left( 1- \frac{\gammam}{\sqrt{2 \Omega_{\textrm{N}} \Omegas}} \right) + \frac{2\Omega_{\textrm{N}}}{\Omegas} - 2^{3/2} \gammam \sqrt{\frac{\Omega_{\textrm{N}}}{\Omegas^3}} \\[1.5mm] i \left( 1- \frac{\gammam}{\sqrt{2 \Omega_{\textrm{N}} \Omegas}} \right) +
%        \frac{2\Omega_{\textrm{N}}}{\Omegas}  - 2^{3/2} \gammam \sqrt{\frac{\Omega_{\textrm{N}}}{\Omegas^3}} & \frac{\Omega_{\textrm{N}}^2}{\gammam \omegam} - \frac{2^{3/2}}{\omegam}\sqrt{\frac{\Omega_{\textrm{N}}^3}{\Omegas}} + \frac{\gammam}{\omegam}\frac{4\Omega_{\textrm{N}}}{\Omegas} + \frac{\gammam^2}{\omegam}\frac{\Omegas^2 - 16\Omega_{\textrm{N}}^2}{2^{5/2} (\Omega_{\textrm{N}} \Omegas)^{3/2}}
%    \end{bmatrix}
%\end{align}}}
%\end{widetext}
{
\begin{align}
\widetilde{\mathbf{V}}^{bv}_{\text{pt}} \dotproduct  \mathbf{V}_M \dotproduct
(\widetilde{\mathbf{V}}_\text{pt}^{bv})^T \approx \widetilde{\mathbf{V}}^{bb} + \begin{bmatrix}
         2^{3/2} \omegam \sqrt{\frac{\Omega_{\textrm{N}}}{\Omegas^3}} & -i + \frac{2\Omega_{\textrm{N}}}{\Omegas} \\[1.5mm] i +
        \frac{2\Omega_{\textrm{N}}}{\Omegas} & - \frac{2^{3/2}}{\omegam}\sqrt{\frac{\Omega_{\textrm{N}}^3}{\Omegas}}
    \end{bmatrix}
\end{align}}
Therefore, Eq. (\ref{eqn:entanglement_indicator}) reduces to the following condition in the limit of $\gammam \rightarrow 0$:
\begin{align}
\label{eq:entanglement_threshold_white_noise}
    \Omega_{\textrm{N}} < \Omegas \iff \text{optomechanical entanglement}.
\end{align}
When $\eta = 1$, i.e. there are no passive losses, this relation gets modified as $\Omegaf < \Omegas \iff \text{entanglement}$, which signifies that the presence of optomechanical entanglement depends on a simple trade-off between the relative strengths of the environmental noise sources. This bound has been observed numerically in \cite{direkci_2024}. For $\eta < 1$, $\Omega_{\textrm{N}} > \Omegaf$, therefore it becomes harder to achieve optomechanical entanglement for non-unity photodetector efficiency, and it is more advantageous to have a small coherent coupling $\Omq$ for this purpose.

We compute the logarithmic negativity numerically, assuming that the input light field is the vacuum state, and the environmental noise sources are white, characterized by $\Omegaf$ and $\Omegas$ in Eq. (\ref{eq:white_force_sensing}). The results are given in Fig. \ref{fig:dark_noise}. In order to work in the free-mass limit, we fix $\omegam/(2\pi) = 1$ Hz, $\gammam/(2\pi) = 0.01$ Hz, $\Omegaf/(2\pi) = 10$ Hz, and vary $\Omegas$ in the range of $\Omegas/(2\pi) \in [1,\, 500]$ Hz, such that $\Omegaf, \Omegas \gg \omegam, \gammam$.
First, we assume a perfect photodetector with unity efficiency ($\eta = 1$). This case is plotted with dashed and plain lines in blue, where the interaction strength is taken to be $\Omq/(2\pi) = 10$ Hz, and $\Omq/(2\pi) = 20$ Hz, respectively. The entangling threshold occurs at $\Omegaf \approx \Omegas$, as observed in Fig. \ref{fig:dark_noise}. Then, we assume a photodetector with non-unity efficiency ($\eta = 0.9$), which introduces passive losses to the system. We plot this case with dashed and plain lines in yellow, for $\Omq/(2\pi) = 10$ Hz, and $\Omq/(2\pi) = 20$ Hz, respectively. 
From Eq. (\ref{eq:entanglement_threshold_white_noise}), we compute the entangling threshold as $\Omegas/\Omegaf = 1.025$ ($\Omegas/\Omegaf = 1.095$) for $\Omq/(2\pi) = 10$ Hz ($\Omq/(2\pi) = 20$ Hz), which is drawn to the inset of Fig. \ref{fig:dark_noise} with a red, vertical curve. As observed analytically, passive losses impede achieving optomechanical entanglement. Furthermore, it is more advantageous to have a small interaction strength to obtain entanglement (i.e. reach the entangling threshold). However, once the threshold is reached, a larger interaction strength provides more entanglement, quantified by a larger logarithmic negativity.

\section{Generalization to arbitrary squeezing}
\label{sec:fd_squeezing}

Until this part of the paper, we have assumed that the input light field consisted of vacuum fluctuations. Now, let us assume that arbitrary squeezing operations have been performed on the input field before interacting with the mechanical oscillator.

First, we show in Appendix \ref{app:squeezing_or_rotations} that if the input field has been subjected to phase shifts or squeezing by a constant factor, the determinant condition in Eq. (\ref{eqn:entanglement_indicator}) is unchanged. Hence, such operations do not affect the entangling-disentangling transition. Therefore, we must apply a more general, potentially frequency-dependent transformation to the input light field in order to act on the entangling-disentangling transition.

%A general transformation performed on the light field must preserve the commutation relations, hence, it must be symplectic. 
A general, unitary transformation performed on the input field must be causal and symplectic. The causality condition is because we have defined the input field to be the temporal modes of the light during $t>0$. Then, any non-causal transformation mixes the input and the output temporal modes, which conflicts with the causality of the dynamics. 

The most general causal symplectic transformation can be written as (see Appendix \ref{app:general_symplectic_transformations})
\begin{align}
    \begin{bmatrix}
        \hat u_1(\Omega) \\ \hat u_2(\Omega)
    \end{bmatrix} \rightarrow e^{i \phi(\Omega)} \left[\begin{array}{cc}
      A(\Omega) & B(\Omega) \\
     C(\Omega) & D(\Omega)
\end{array}\right] \begin{bmatrix}
        \hat u_1(\Omega) \\ \hat u_2(\Omega)
    \end{bmatrix}
\end{align}
where $A(\Omega)$, $B(\Omega)$, $C(\Omega)$, and $D(\Omega)$ are real functions of $\Omega$, and $A(\Omega)D(\Omega)-B(\Omega)C(\Omega)= 1\,\forall \,\Omega$. The phase factor $e^{i \phi(\Omega)}$ ensures the causality of the transformation. 

We proceed with writing down the covariance matrix of the system, given this transformation. First, we calculate the correlations between the quadratures of the output light field,
\begin{subequations}
\label{eqn:fd_vv_cov}
\begin{align}
    V^{vv}_{11}(\Omega) &= T_1(\Omega) \label{eqn:fd_vv_cov_v11} \\
    V^{vv}_{12}(\Omega) &= T_{12}(\Omega) + \Omq^2 \, T_1(\Omega) \chi(-\Omega) \label{eqn:fd_vv_cov_v12} \\
    V^{vv}_{22}(\Omega) &= T_2(\Omega) + \Omq^4 \, T_1(\Omega) |\chi(\Omega)|^2  + \Omq^2 \left[ \chi_\text{EN}(\Omega) \right. \nonumber \\  & \left.+ T_{12}(\Omega) \cdot (\chi(\Omega)+\chi(-\Omega)) \right] \label{eqn:fd_vv_cov_v22}
\end{align}
\end{subequations}
where we have defined \mbox{$T_1(\Omega) = A(\Omega)^2 + B(\Omega)^2$}, $T_2(\Omega) = C(\Omega)^2 + D(\Omega)^2$, and \mbox{$T_{12}(\Omega) = A(\Omega)C(\Omega) + B(\Omega) D(\Omega)$}.
Next, the covariances of the quadratures of the mechanical oscillator are given by
\begin{align}
\label{eqn:fd_bb_cov}
    V^{bb}_{11}(\Omega) = \omegam\chiF(\Omega) + \omegam\Omq^2 |\chi(\Omega)|^2 T_1(\Omega),
\end{align}
and $V^{bb}_{12}(\Omega) = i\Omega V^{bb}_{11}(\Omega)/\omegam, \; V^{bb}_{22}(\Omega) = \Omega^2 V^{bb}_{11}(\Omega)/\omegam$. We see that the variance associated with the force noise is unchanged compared to Eq. (\ref{eqn:mech_cov_matrix}), and the variance due to the optomechanical interaction is scaled by the frequency-dependent factor of $T_1(\Omega)$ in the frequency domain. Finally, we compute the cross-spectra of the mechanical oscillator and the output light field:
\begin{subequations}
\label{eqn:fd_cov_light_mech}
\begin{align}
V^{bv}_{11}(\Omega)
     &= \Omq \sqrt{\omegam}\chi(\Omega) T_1(\Omega),  \label{eqn:fd_cov_light_mech_v1} \\
     V^{bv}_{12}(\Omega)
     &=  \Omq \sqrt{\omegam} T_{12}(\Omega) \chi(\Omega) + \Omq \sqrt{\omegam} \chiF(\Omega) \nonumber \\ &+ \Omq^3 \sqrt{\omegam} |\chi(\Omega)|^2 T_1(\Omega) \label{eqn:fd_cov_light_mech_v2}
\end{align}
\end{subequations}
and $V^{bv}_{2i}(\Omega) = -i \Omega \, V^{bv}_{1i}(\Omega) /\omegam$, $i = 1, 2$. To continue with the formalism of Sec. \ref{sec:universal_entanglement}, let us recall the general form of the entanglement criterion. From Eq. (\ref{eqn:determinant_full}) and (\ref{eqn:breaking_down_big_matrix_general}), we have:
\begin{align}
\label{eqn:det_full_general_fd}
    &\mathrm{det} \left[ \mathbf{V}^{bb} + i \mathbf{K}^b - \mathbf{V}^{bv}_{\text{\text{pt}}} \boldsymbol{\cdot} \left(\mathbf{\Sigma} \dotproduct \mathbf{D}_M \dotproduct \mathbf{\Sigma}^\dagger  + \mathbf{\Delta} \right) \boldsymbol{\cdot} (\mathbf{V}^{bv}_{\text{\text{pt}}})^T \right] < 0 \nonumber \\ &\quad \iff \text{entanglement}.
\end{align}
where $\mathbf{\Delta}$, $\mathbf{\Sigma}$, and $\mathbf{D}_M$ are given as (see Eq. (\ref{eqn:inverse_schur}))
\begin{subequations}
\label{eqn:fd_matrix_defns}
\begin{align}
&\mathbf{\Delta} = \left[ \begin{array}{cc}
 V^{vv}_{11}(\tau)^{-1}  & 0 \\
 0 & 0  \end{array} \right], \\
&\mathbf{\Sigma} = \begin{bmatrix}
     V^{vv}_{11}(\tau)^{-1} \dotproduct V^{vv}_{12}(\tau) + i V^{vv}_{11}(\tau)^{-1} & 0 \\
 -\delta(\tau) & 0
\end{bmatrix}, \label{eq:sigma_defn} \\
& \mathbf{D}_M = \begin{bmatrix}
    M^{-1}(\tau) & 0 \\
    0 & M^{-1}(\tau)
\end{bmatrix}.
\end{align}
\end{subequations}
and $\mathbf{V}^{bv}_{\text{\text{pt}}}$ is defined in Eq. (\ref{eqn:v_pt_general}). To compute Eq. (\ref{eqn:det_full_general_fd}), we need to invert $V^{vv}_{11}$ using the Wiener-Hopf method, therefore, we write it down in the Fourier domain as a product of its causal and anti-causal poles and zeros,
\begin{align}
\label{eq:V^vv_11_wiener_hopf}
    V^{vv}_{11}(\Omega) = T_1(\Omega) = T_1(\Omega)_+T_1(\Omega)_-,
\end{align}
where $T_1(\Omega)_+^\dagger = T_1(\Omega)_-$, and $T_1(\Omega)_+$ ($T_1(\Omega)_-$) contains the causal (anti-causal) poles and zeros of $V^{vv}_{11}(\Omega)$.
In the absence of any squeezing in the input light field, we had found a simplification in the entanglement criterion, which we referred to as vacuum cancellation in Sec. \ref{sec:entanglement_indicator}. This simplification canceled out some of the terms in the entanglement criterion related to the mechanical oscillator; cf. Eq.~(\ref{eq:determinant_after_vacuum_cancellation}). Specifically, we observed that the remaining terms were the variances of $\hat b_1$ and $\hat b_2$ generated due to the force noise only. Here, in the presence of frequency-dependent squeezing, we observe that the vacuum cancellation persists. To see this, we operate on $\mathbf{V}^{bv}_{\text{pt}}$ with $\mathbf{\Delta}$ from the right,
\begin{align}
    (\mathbf{V}^{bv}_{\text{pt}} \dotproduct \mathbf{\Delta})(\Omega) = \begin{bmatrix}
    \;\;\, V^{bv}_{11} \dotproduct (V^{vv}_{11})^{-1} & 0 \\[1mm] -V^{bv}_{21} \dotproduct (V^{vv}_{11})^{-1} & 0
\end{bmatrix}.
\end{align}
Using Eq. (\ref{eqn:fd_cov_light_mech}) and Eq. (\ref{eq:V^vv_11_wiener_hopf}), we can simplify this expression to obtain
\begin{align}
\label{eqn:fd_vacuum_cancellation}
&\sqrt{\omegam}\Omq \, 
\frac{1}{T_1(\Omega)_+} \left[\frac{T_1(\Omega)_+ T_1(\Omega)_- \chi(\Omega)}{T_1(\Omega)_-}  \right]_+ \begin{bmatrix}
     1 & 0 \\ -\frac{i\Omega}{\omegam} & 0
 \end{bmatrix} \nonumber \\ &= \sqrt{\omegam}\Omq \, \chi(\Omega) \begin{bmatrix}
     1 & 0 \\ -\frac{i\Omega}{\omegam} & 0 
 \end{bmatrix}
\end{align}
Then, in the frequency domain, we have
\begin{align}
&\left( \mathbf{V}^{bv}_{\text{pt}} \dotproduct \mathbf{\Delta} \dotproduct (\mathbf{V}^{bv}_{\text{pt}})^T \right)= \omegam\Omq^2 \, |\chi(\Omega)|^2 T_1(\Omega) \begin{bmatrix}
     1 & \frac{i\Omega}{\omegam} \\[1.5mm] -\frac{i\Omega}{\omegam} & \frac{\Omega^2}{\omegam^2}
 \end{bmatrix}
\end{align}
This term is equal (in the frequency domain) to the variances of the quadratures of the mechanical oscillator arising from the optomechanical interaction, cf. the second term in Eq. (\ref{eqn:fd_bb_cov}). Therefore, we see that the vacuum cancellation is realized, and we can use Eq. (\ref{eq:determinant_after_vacuum_cancellation}) to test for entanglement, given that $\mathbf{\Sigma}$ and $\mathbf{D}_M$ are calculated accordingly.

Continuing with the calculation, we want to compute $\mathbf{V}^{bv}_{\text{pt}} \dotproduct \mathbf{\Sigma} \dotproduct \mathbf{D}_M \dotproduct \mathbf{\Sigma}^\dagger  \dotproduct ({\mathbf{V}^{bv}_{\text{pt}}})^T$. For this purpose, we operate on $\mathbf{V}^{bv}_{\text{pt}}$ from the right with $\mathbf{\Sigma}$. Using Eq. (\ref{eqn:v_pt_general}) and Eq. (\ref{eq:sigma_defn}), we obtain 
\begin{align}
\begin{bmatrix}
      \;\;\, V^{bv}_{11}(t) \dotproduct (V^{vv}_{11})^{-1} \dotproduct (i\delta(\tau) + V^{vv}_{12}(\tau)) - V^{bv}_{12}(t) & 0 \\[0.8mm] - V^{bv}_{21}(t) \dotproduct (V^{vv}_{11})^{-1} \dotproduct (i\delta(\tau) + V^{vv}_{12}(\tau)) + V^{bv}_{22}(t) & 0
    \end{bmatrix}.\label{eq:v_m_prod_v_vb_simplifications}
\end{align}
We can simplify this expression by plugging in what we have found for $V^{bv}_{i1}(t) \dotproduct (V^{vv}_{11})^{-1}$, $i=1,2$, from Eq. (\ref{eqn:fd_vacuum_cancellation}), which results in 
\begin{align}
\label{eq:v_m_prod_v_vb_simplifications_2}
    \begin{bmatrix}
      \;\;\, \sqrt{\omegam} \Omq \chi(t) \dotproduct (i\delta(\tau) + V^{vv}_{12}(\tau)) - V^{bv}_{12}(t) & 0 \\[0.8mm] - \frac{\Omq}{\sqrt{\omegam}} \dot \chi(t) \dotproduct (i\delta(\tau) + V^{vv}_{12}(\tau)) + V^{bv}_{22}(t) & 0
    \end{bmatrix}.
\end{align}
To further simplify this expression, we want compute $\chi(t) \dotproduct V^{vv}_{12}(\tau)$. Then, in the frequency domain, we have
\begin{flalign}
    & \Omq \sqrt{\omegam}\,\mathcal{F} \left\{ \int_0^\infty dt' V^{vv}_{12}(t-t') \chi(t')\right\} \nonumber\\
    &= \Omq \sqrt{\omegam}\,\mathcal{F} \left\{ \int_{-\infty}^\infty dt' V^{vv}_{12}(t-t') \chi(t')\right\} \nonumber\\
    &= \Omq \sqrt{\omegam} T_{12}(\Omega) \chi(\Omega) + \Omq^3 \sqrt{\omegam}|\chi(\Omega)|^2 T_1(\Omega) \nonumber \\
    &= V^{bv}_{12} \rvert_{\chiF = 0}
\end{flalign}
where $V^{bv}_{12}$ is given in Eq. (\ref{eqn:fd_cov_light_mech_v2}). Similarly,
\begin{align}
    \frac{\Omq}{\sqrt{\omegam}} \,\mathcal{F} \left\{ \int_0^\infty dt' V^{vv}_{12}(t-t') \dot \chi(t')\right\} = V^{bv}_{22} \rvert_{\chiF = 0}.
\end{align}
Then, similar to the case in Sec. \ref{sec:entanglement_indicator}, the terms related to the optomechanical interaction in $V^{bv}_{12}$ and $V^{bv}_{22}$ cancel out of Eq. (\ref{eq:v_m_prod_v_vb_simplifications_2}).

We realize that thanks to these simplifications, the entanglement criterion in Eq. (\ref{eqn:det_full_general_fd}) is equivalent to the entanglement indicator defined in Eq. (\ref{eqn:entanglement_indicator}). However, while using Eq. (\ref{eqn:entanglement_indicator}), the operator $M$ needs to be recomputed using Eq. (\ref{eq:m_defn_general}). The expressions for the rest of the terms in Eq. (\ref{eqn:entanglement_indicator}), meaning $\widetilde{\mathbf{V}}^{bb}$, $\widetilde{\mathbf{V}}^{bv}_{\text{pt}}$, and, $\mathbf{V}_M$, are unchanged compared to the case where the input field is the vacuum state.

Now, let us calculate $M(t, t')$, whose definition is given in Eq. (\ref{eq:m_defn_general}). Different from Sec. \ref{sec:universal_entanglement}, this operator has two time indices in the presence of arbitrary frequency-dependent squeezing. From the Wiener-Hopf method, we can first compute the columns of the matrix $V^{vv}_{11}(t, t')^{-1} \dotproduct \left[i \delta(t, t') + V^{vv}_{12}(t, t') \right]$. These columns will be functions of $t>0$. Let us refer to them as $\{c_{t'}(t)\}$, where $t'>0$ represents the column index, and $t$ represents the row index, respectively. From Eqs. \cref{eqn:fd_vv_cov_v11,eqn:fd_vv_cov_v12}, we have in the frequency domain
\begin{align}
    c_{t'}(\Omega) &= \frac{1}{T_1(\Omega)_+} \left[ \mathcal{F}\{i \delta(t) + V^{vv}_{12}(-t) \}  \frac{ e^{i \Omega t'}}{T_1(\Omega)_-} \right]_+ \nonumber \\ &= \frac{1}{T_1(\Omega)_+} \left[ \frac{e^{i \Omega t'}}{T_1(\Omega)_-} \left(i   +  T_{12}(\Omega) \right) \right]_+ \nonumber \\ & \quad + \Omq^2\, e^{i \Omega t'} \, \chi(\Omega).
\end{align}
Then, the elements of the matrix $\left[-i \delta(t, t') + V^{vv}_{21}(t, t')\right] \dotproduct V^{vv}_{11}(t, t')^{-1} \dotproduct \left[i \delta(t, t') + V^{vv}_{12}(t, t') \right]$ are given by
\begin{subequations}
\label{eqn:fd_3_toeplitz}
\begin{align}
&\int_{0}^{\infty} dx \left(-i \delta(t-x) + V^{vv}_{21}(x-t) \right) c_{t'}(x), \\
& \int_{-\infty}^{\infty} dx \left(-i \delta(t-x) + V^{vv}_{21}(x-t) \right) c_{t'}(x), \label{eqn:fd_3_toeplitz_b} \\
& \int_{-\infty}^{\infty} \frac{d\Omega}{2\pi} \mathcal{F} \{-i \delta(t-x) + V^{vv}_{21}(x-t) \} c_{t'}(\Omega). \label{eqn:fd_3_toeplitz_c}
\end{align}
\end{subequations}
We were able to extend the lower integration limit to $-\infty$ in Eq. (\ref{eqn:fd_3_toeplitz_b}) since the function $c_{t'}(t)$ is causal. Then, we used the Plancherel theorem to go from the time domain to the frequency domain in Eq. (\ref{eqn:fd_3_toeplitz_c}).
We can write Eq. (\ref{eqn:fd_3_toeplitz_c}) explicitly as
\begin{align}
\label{eqn:fd_3_toeplitz_full}
    & \int_{-\infty}^{\infty} \frac{d\Omega}{2\pi} \biggl\{ \frac{-i + T_{12}(\Omega) }{T_1(\Omega)_+} e^{-i \Omega t} \left[  \frac{e^{i \Omega t'}}{T_1(\Omega)_-} 
(i+ \, T_{12}(\Omega)) \right]_+   \nonumber \\ &+ \Omq^2\, T_{12}(\Omega) e^{i \Omega (t'-t)} \, (\chi(\Omega) + \chi(-\Omega)) -i \Omq^2\, e^{i \Omega (t'-t)} \nonumber \\ &  (\chi(\Omega)-\chi(-\Omega)) + \Omq^4 \, e^{i \Omega (t'-t)} |\chi(\Omega)|^2 T_{1}(\Omega) \biggr\}.
\end{align}
This expression describes the elements of a matrix indexed by $t, t'>0$, where $t$ and $t'$ are the row and the column indices, respectively.
We observe that this matrix is Hermitian, but not necessarily Toeplitz. The Hermitian property is seen from the fact that interchanging $t$ and $t'$ corresponds to the conjugation of this expression. However, this matrix can be written as a sum of a Hermitian matrix and a Hermitian Toeplitz matrix,
where the Toeplitz part is the sum of the terms in Eq. (\ref{eqn:fd_3_toeplitz_full}) that are a function of $t-t'$ (i.e. the terms in 2nd and 3rd row). 

Therefore, $M$ will also be the sum of a Hermitian matrix and a Hermitian Toeplitz matrix. More specifically, using Eqs. (\ref{eq:m_defn_general}), (\ref{eqn:fd_vv_cov_v22}), and (\ref{eqn:fd_3_toeplitz_full}), we find that $M$ can be written as
\begin{align}
    M(t,t') = M_{\text{vac}}(t-t') + \Lambda(t, t')
\end{align}
where $M_{\text{vac}}(t-t')$ is what we have found for $M$ when the input light field is the vacuum state (given in Eq. (\ref{eqn:m_vac_defn})), and $\Lambda(t,t')$ is some Hermitian operator independent of $\Omq$, whose entries depend on the squeezing in the input light field. The full expression for $\Lambda(t,t')$ is written as
\begin{align}
\label{eq:lambda_fd_defn}
    \Lambda(t,t') = T_2(t,t') -&\int_{-\infty}^{\infty} \frac{d\Omega}{2\pi} \left( \frac{-i + T_{12}(\Omega) }{T_1(\Omega)_+} \right) e^{-i \Omega t} \cdot \nonumber \\
    &\left[  \frac{e^{i \Omega t'}}{T_1(\Omega)_-} 
(i+ \, T_{12}(\Omega)) \right]_+.
\end{align}
We observe that when the input light field is the vacuum state, $T_2(t,t') = \delta(t-t')$, $T_{12}(\Omega) = 0$, and $T_1(\Omega)_\pm = 1$. Then, using Eq. (\ref{eq:lambda_fd_defn}), we find that
$\Lambda = \mathbf{0}$ for this case, and $M(t,t') = M_{\text{vac}}(t-t')$.

We want to investigate whether optomechanical entanglement is achievable with arbitrary, frequency-dependent squeezing in the input light field. For this purpose, we first observe that when $\Omq \rightarrow \infty$, $M \rightarrow M_{\text{vac}}$, and the determinant condition in Eq. (\ref{eqn:entanglement_indicator}) is equivalent to its form when the input light field is the vacuum state for a given force and sensing noise. We refer to this configuration as the ``vacuum case". 

Now, we want to show that if the system is entangled for some $\Omq$, it is bound to be entangled for $\Omq' > \Omq$. 
%In this case, since the $\Omq \rightarrow \infty$ limit is equivalent to the vacuum case, if optomechanical entanglement exists for some frequency-dependent squeezing configuration, it is bound to exist for the vacuum case (assuming that the force and sensing noise spectra are unchanged). In other words, the hyperspace of force and sensing noise spectra for which there exists optomechanical entanglement for the vacuum case encapsulates the space when the input light field is squeezed arbitrarily.
To show this, let us assume that the system is entangled for some $\Omq$. We notice that the first two terms in Eq. (\ref{eqn:entanglement_indicator}) will be independent of $\Omq$. The $\Omq$ dependency comes from the third term, where $ \widetilde{\mathbf{V}}^{bv}_{\text{pt}} \propto \Omq$, and  $M(t,t') = M_{\text{vac}}(t-t') + \Lambda(t, t')$, $M_{\text{vac}} \propto \Omq^2$. Then, the third term can be written in the form of 
\begin{align}
\label{eq:fd_final_proof_1}
    &\mathbf{C} \dotproduct \begin{bmatrix}
        \left( \frac{1}{\Omq^2}\Lambda + D \right)^{-1} & 0 \\
        0 & \left( \frac{1}{\Omq^2}\Lambda + D \right)^{-1}
    \end{bmatrix} \dotproduct \mathbf{C}^\dagger, \nonumber \\ &\mathbf{C} = \frac{1}{\Omq} \widetilde{\mathbf{V}}^{bv}_{\text{pt}} \dotproduct \begin{bmatrix}
    -i \delta(\tau) & 0 \\ \delta(\tau) & 0
\end{bmatrix}, \;\; D = \frac{1}{\Omq^2} M_{\text{vac}}
\end{align}
where $\mathbf{C}$, $D$, and $\Lambda$ are independent of $\Omq$. Furthermore, notice that $M$ is the Schur complement of the block $V^{vv}_{11}$ of the matrix $\mathbf{V}^{vv} + i\mathbf{K}^v$. Since $V^{vv}_{11}$ is positive definite, and $\mathbf{V}^{vv} + i\mathbf{K}^v$ is positive semi-definite (from the Heisenberg uncertainty principle), $M$ is positive semi-definite for all $\Omq$. Consequently, $\Lambda$ and $M_{\text{vac}}$ are also positive semi-definite, since $M = \Lambda$ when $\Omq = 0$, and $M \rightarrow M_{\text{vac}}$ in the limit of $\Omq \rightarrow \infty$. Also, recall that both $D$ and $\Lambda$ are Hermitian. As a result, we can write for $\Omq'>\Omq$
\begin{align}
\label{eq:fd_psd_inequality}
    \frac{1}{{\Omq'}^2} \Lambda &< \frac{1}{{\Omq}^2} \Lambda, \nonumber \\
    \frac{1}{{\Omq'}^2} \Lambda + D &< \frac{1}{{\Omq}^2} \Lambda + D, \nonumber \\
    \left( \frac{1}{{\Omq'}^2} \Lambda + D \right)^{-1} &> \left( \frac{1}{{\Omq}^2} \Lambda + D \right)^{-1}.
\end{align}
Therefore, if we refer to the matrix in the determinant condition in Eq. (\ref{eqn:entanglement_indicator}) as $\mathbf{V}_\text{det}(\Omq)$, we can write $\mathbf{V}_\text{det}(\Omq') < \mathbf{V}_\text{det}(\Omq)$ for $\Omq' > \Omq$. Since the system is entangled for $\Omq$, the determinant of $\mathbf{V}_\text{det}(\Omq)$ is negative. Furthermore, since we are looking for the entanglement between one mode and the rest, $\mathbf{V}_\text{det}(\Omq)$ can have at most one negative eigenvalue (for this case, it has exactly one negative eigenvalue). Then, $\mathbf{V}_\text{det}(\Omq')$ will also have a negative eigenvalue, and the system will be entangled.
Hence, we proved that if optomechanical entanglement exists for $\Omq$, it also exists for $\Omq'>\Omq$. 

\begin{figure}
    \centering
\includegraphics[width=\linewidth]{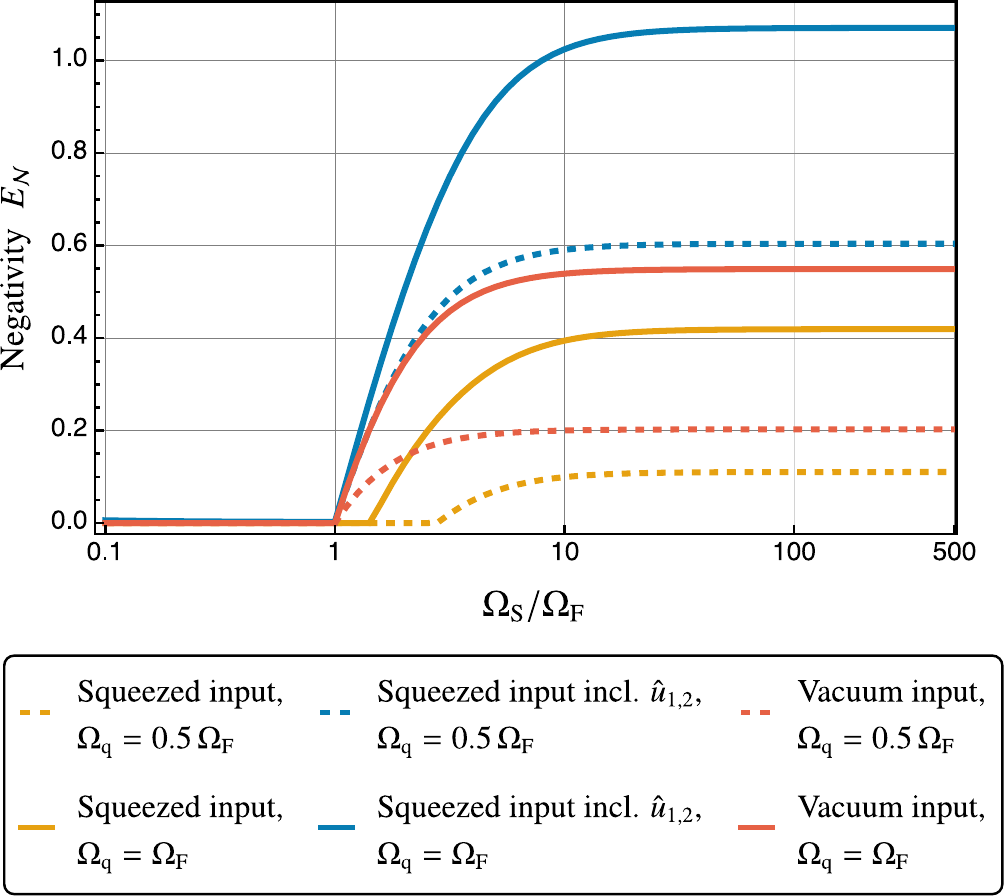}
\caption{Logarithmic negativity for the case where the input light field is frequency-dependently squeezed with $r=1$ (c.f. Eq. (\ref{eq:quantum_noise_squeezing})), and for the vacuum input. We further consider the cases where we compute the entanglement between the mechanical oscillator at time $t=0$ and the light field, where the field is considered to be i) the outgoing field that has left the oscillator during $t<0$, ii) the joint field combining the outgoing field that has left the oscillator during $t<0$, and the ingoing field that has yet to interact with the oscillator, but will be reflected from the oscillator during $t>0$. These cases (for a squeezed input) are plotted in yellow and blue curves, respectively. For the vacuum input plotted in red, we only consider case i). We observe that for a squeezed input, entanglement is harder to achieve for case i), and the entangling transition is not universal with respect to the interaction strength $\Omq$. However, for case ii), universality is restored.}
\label{fig:freq_dependent_squeezing}
\end{figure}

Consequently, we can make the following statements about the nature of optomechanical entanglement when the input light field is squeezed in a frequency-dependent fashion. First, we observe that universality with respect to the interaction strength, $\Omq$, does not hold, since the entanglement indicator in Eq. (\ref{eqn:entanglement_indicator}) depends on $\Omq$. Hence, increasing $\Omq$ may provide optomechanical entanglement to form. However, the transition is still unique with respect to the amount of environmental noise. Second, if there exists optomechanical entanglement for any frequency-dependent squeezing configuration, it will also exist when the input light field is in the vacuum state. Therefore, it is ``easier" to achieve entanglement for a given force and sensing noise without frequency-dependently squeezing the input light field. In other words, the hyperspace of force and sensing noise spectra for which there exists optomechanical entanglement for the vacuum case encapsulates the space when the input light field is squeezed arbitrarily \footnote{Note that the space is unchanged for frequency-independent squeezing.}. We observed numerically that after optomechanical entanglement is achieved, the amount of entanglement, quantified by the logarithmic negativity, can be greater or less than the case where the input light field is the vacuum state, depending on the frequency-dependent squeezing transformation.

\section{Discussion for Frequency-Dependent Squeezing}

In Sec. \ref{sec:fd_squeezing}, it was proven that when the input light field is squeezed in a frequency-dependent fashion, it is harder to achieve optomechanical entanglement compared to the vacuum input case, i.e. there  exists a region of the parameter space where optomechanical entanglement is observed only for the latter configuration. Intuitively, this result can be understood in the following way: when the ingoing light field is the vacuum state, the modes corresponding to different frequencies (or arrival times with respect to the surface of the oscillator) are not correlated. Conversely, when the ingoing light field is frequency-dependently squeezed, there is non-zero correlation between these modes. Then, when we consider the outgoing light field that has reflected from the mechanical oscillator during time $t<0$, it will be entangled with the ingoing light field that has yet to interact with the oscillator during this time (note that this field will be reflected from the oscillator during time $t>0$). Hence, while computing the entanglement between this field and the mechanical oscillator at time $t=0$, the correlation will induce an additional noise to the system when the ingoing light modes are traced out. Therefore, we can expect that it is harder to get entangled for this case, as it will appear to be more noisy than the vacuum input case.

This extra noise can be overcome by computing the entanglement between the mechanical oscillator at time $t=0$, and the joint field combining the outgoing field that has left the oscillator during $t<0$, and the ingoing field that has yet to interact with the oscillator. Since the joint field contains information about the light field during time $t \in (-\infty, \infty)$, the frequency-dependent squeezing can be ``reversed" \footnote{The squeezing cannot be reversed perfectly due to the interaction with the mechanical oscillator.} by whitening the joint field. This would not be possible if we had only access to the outgoing light field, as the whitening transformation will be non-causal in the presence of frequency-dependent squeezing.

This situation is illustrated in Fig. \ref{fig:freq_dependent_squeezing}. We fix the parameters of the mechanical oscillator as $\omegam/(2\pi) = 1$ Hz, $\gammam/(2\pi) = 0.01$ Hz, and assume white environmental noise sources parametrized by $\Omegaf/(2\pi) = 10$ Hz, and $\Omegas/(2\pi) \in [1, 500]$ Hz. We first plot the logarithmic negativity for the vacuum input with red curves, for interaction strengths of $\Omq/(2\pi) = 5$ Hz and $\Omq/(2\pi) = 10$ Hz, with dashed and plain curves, respectively. We repeat this calculation for a frequency-dependently squeezed ingoing field (plotted in yellow). The squeezing transformation is such that the total quantum noise is squeezed with
\begin{align}
\label{eq:quantum_noise_squeezing}
    S(\Omega) \rightarrow e^{-2r} S(\Omega),
\end{align}
where $S(\Omega)$ is the spectrum of the quantum noise, and $r$ is a squeezing parameter. For more details about the realization, see Appendix \ref{app:freq_dep_squeezing}. We fix $r=1$, and observe that the entangling transition is not universal with respect to the interaction strength $\Omq$ in the presence of frequency-dependent squeezing, as was shown in Sec. \ref{sec:fd_squeezing}. Finally, we repeat the calculation for the same squeezed input, but we compute the entanglement between the oscillator and the joint field combining the outgoing and the ingoing light field during $t<0$ (plotted in blue). We observe that the universality is restored, and the entangling transition occurs at $\Omegaf \approx \Omegas$, as found in Sec. \ref{sec:example}. Note that the amount of entanglement is the largest for this case.

%%%%%%%%%%%%%%%%%%%%%%%%%%%%%%%%%%%%%%%%%%%%%%%%%%%%%%%%%%%%%%%%%%%%%%%%%%%%%

\section{Conclusions}

In this paper, we proved the existence of an entangling-disentangling transition in an optomechanical system consisting of a mechanical oscillator interacting with a light field in a pure state. We assumed arbitrary, possibly non-Markovian, environmental noise sources. Assuming Gaussian dynamics, we were able to compute necessary and sufficient conditions for entanglement using the PPT criterion.

First, assuming that the input light field is the vacuum state, we showed that the entangling-disentangling transition is independent of the interaction strength between the oscillator and the light field, hence, it is universal. Therefore, if the environmental noises are such that the system is not in the regime where optomechanical entanglement can occur, one cannot achieve it by increasing the interaction strength between the oscillator and the light field. 

In the second part of the paper, we assumed arbitrary Gaussian and causal unitary transformations applied to the light field, including frequency-dependent squeezing. We demonstrated that the entangling-disentangling transition is not affected by frequency-independent squeezing and phase shifts -- i.e. it is equivalent to the vacuum input case. However, when the light field is squeezed in a frequency-dependent manner, we showed that entanglement is harder to achieve. More specifically, we proved that if there is no optomechanical entanglement when the input light field is the vacuum state, it is impossible to generate it by squeezing the input light field arbitrarily.

We expand on the physical interpretation and consequences of these results in a companion Letter \footnotemark[1].

\begin{acknowledgments}

S.D. and Y.C. acknowledge the support by the Simons Foundation (Award Number 568762), and by the US NSF
grant PHY-2309231. K.W. acknowledges the
support by the Vienna Doctoral School in Physics (VDSP). K.W., C.G., and M.A. received
funding from the European Research Council (ERC) under the
European Union’s Horizon 2020 research and innovation program (Grant Agreement No. 951234), and from the Research
Network Quantum Aspects of Spacetime (TURIS).

\end{acknowledgments}

%%%%%%%%%%%%%%%%%%%%%%%%%%%%%%%%%%%%%%%%%%%%%%%%%%%%%%%%%%%%%%%%%%%%%%%%%%%%%%%%%%%%%%%%%%%%%%%%%%%%%%%%%%%%%%%

\appendix

\section{Equations of motion with physical dimensions}
\label{app:equations_of_motion}

Here, we present the equations of motion with physical dimensions. We denote the position (momentum) of the center of mass of the mechanical oscillator with $\hat{x}$ ($\hat{p}$), and the amplitude and phase quadratures of the input (output) light field with $\hat{u}_1(t), \,\hat{u}_2(t)$($\hat{v}_1(t), \,\hat{v}_2(t)$), respectively. The equations read
\begin{subequations}
\begin{align}
\hat{v}_1(t) &= \hat{u}_1(t) \,, \\
\hat{v}_2(t) &= \hat{u}_2(t) + \alpha (\hat{x}(t)+\hat{N}_{\rm{S}}(t)) \,, \\
\frac{d \hat p(t)}{dt} &= -\gammam \hat{p}(t) -m \omegam^2 \hat{x}(t) +\hbar \alpha \hat{u}_1(t) +\hat{N}_{\rm{F}}(t) \,, \\
\frac{d \hat x(t)}{dt} &= \hat{p}(t)/m \,,
\end{align}
\end{subequations}
where $\alpha = \Omq \sqrt{m/\hbar}$, $\Omq$ is the characteristic interaction strength as in the main text, $\gammam$ is the damping rate of the oscillator, $m$ is the mass of the oscillator, and $\hat{N}_{\rm{F}}(t)$ and $\hat{N}_{\rm{S}}(t)$ are the force and the sensing noise, respectively. The quadratures of the mechanical oscillator in the main text, $\hat b_1(t)$ and $\hat b_2(t)$, are related to $\hat{x}(t)$ and $\hat{p}(t)$ with
\begin{align}
    \hat b_1(t) = \sqrt{\frac{2 m \omegam}{\hbar}} \hat x(t)\,, \quad \hat b_2(t) = \sqrt{\frac{2}{\hbar m \omegam}} \hat p(t)
\end{align}
so that $\left[ \hat x(t), \, \hat p(t)\right] = i \hbar$, $\left[ \hat b_1(t), \, \hat b_2(t)\right] = 2i$. Note that here, the commutators are for same-time operators: $\left[ \hat x(t), \, \hat p(t')\right] = 0$ for $t \neq t'$. Furthermore, $\hat{N}_{\rm{F}}(t)$ and $\hat{N}_{\rm{S}}(t)$ are related to $\hat{n}_{\rm{F}}(t)$ and $\hat{n}_{\rm{S}}(t)$ in the main text with
\begin{align}
    \hat N_{\rm{S}}(t) = \sqrt{\frac{\hbar}{2 m \omegam}} \hat n_{\rm{S}}(t), \; \hat N_{\rm{F}}(t) = \sqrt{\frac{\hbar m \omegam}{2}} \hat n_{\rm{F}}(t)
\end{align}
so that $\hat n_{\rm{S}}$ is dimensionless and $\hat n_{\rm{F}}$ has dimensions of Hertz.

\section{Matrix inversion using Schur complement}
\label{app:schur}

In the main text, we encounter block matrices in the form of
\begin{align}
    \mathbf{L} = \begin{bmatrix}
        \mathbf{A} & \mathbf{B} \\
        \mathbf{C} & \mathbf{D}
    \end{bmatrix}
\end{align}
where $\mathbf{A}$, $\mathbf{B}$, $\mathbf{C}$, and $\mathbf{D}$ are square matrices. If $\mathbf{A}$ is invertible, we can write
\begin{align}
    \mathbf{L}^{-1} = \begin{bmatrix}
        \mathbf{A}^{-1} + \mathbf{A}^{-1}\mathbf{B} \left( \mathbf{L}/\mathbf{A} \right)^{-1} \mathbf{C} \mathbf{A}^{-1} & - \mathbf{A}^{-1} \mathbf{B} \left( \mathbf{L}/\mathbf{A} \right)^{-1}\\[1mm]
        - \left( \mathbf{L}/\mathbf{A} \right)^{-1} \mathbf{C} \mathbf{A}^{-1} & \left( \mathbf{L}/\mathbf{A} \right)^{-1}
    \end{bmatrix}
\end{align}
where $\mathbf{L}/\mathbf{A} \coloneqq \mathbf{D} - \mathbf{C} \mathbf{A}^{-1} \mathbf{B}$ is the Schur complement of the block $\mathbf{A}$ of the matrix $\mathbf{L}$. We can restate this formula as
\begin{align}
\label{app_eqn:schur}
    \mathbf{L}^{-1} &= 
    \begin{bmatrix}
        \mathbf{A}^{-1} & \mathbf{0} \\
        \mathbf{0} & \mathbf{1}
    \end{bmatrix} \begin{bmatrix}
        \mathbf{B} \left( \mathbf{L}/\mathbf{A} \right)^{-1} \mathbf{C} & - \mathbf{B} \left( \mathbf{L}/\mathbf{A} \right)^{-1}\\
        - \left( \mathbf{L}/\mathbf{A} \right)^{-1} \mathbf{C} & \left( \mathbf{L}/\mathbf{A} \right)^{-1}
    \end{bmatrix} \begin{bmatrix}
        \mathbf{A}^{-1} & \mathbf{0} \\
        \mathbf{0} & \mathbf{1}
    \end{bmatrix} \nonumber \\
    & \quad +\begin{bmatrix}
        \mathbf{A}^{-1} & \mathbf{0} \\
        \mathbf{0} & \mathbf{0}
    \end{bmatrix} \nonumber \\
    & = \begin{bmatrix}
        \mathbf{A}^{-1} & \mathbf{0} \\
        \mathbf{0} & \mathbf{1}
    \end{bmatrix} \begin{bmatrix}
        \mathbf{B} & \mathbf{0} \\
        -\mathbf{1} & \mathbf{0}
    \end{bmatrix} \begin{bmatrix}
         \left( \mathbf{L}/\mathbf{A} \right)^{-1} & \mathbf{0} \\
        \mathbf{0} & \left( \mathbf{L}/\mathbf{A} \right)^{-1}
    \end{bmatrix} \nonumber \\
    & \quad \, \begin{bmatrix}
        \mathbf{C} & -\mathbf{1} \\
        \mathbf{0} & \mathbf{0}
    \end{bmatrix} \begin{bmatrix}
        \mathbf{A}^{-1} & \mathbf{0} \\
        \mathbf{0} & \mathbf{1}
    \end{bmatrix} + \begin{bmatrix}
        \mathbf{A}^{-1} & \mathbf{0} \\
        \mathbf{0} & \mathbf{0}
    \end{bmatrix}
\end{align}
where $\mathbf{0}$ and $\mathbf{1}$ are the null and the identity matrices, respectively. In the main text, we evaluate $\left(\mathbf{V}^{vv}_{\text{pt}} + i \mathbf{K}^v \right)^{-1}$. Then, we can use the expression derived above, by substituting 
\begin{align}
    \mathbf{A} &= V^{vv}(\tau) \nonumber \\
    \mathbf{B} &= V^{vv}_{12}(\tau) + i \delta(\tau) \nonumber \\
    \mathbf{C} &= \mathbf{B}^\dagger = V^{vv}_{21}(\tau) - i \delta(\tau) \nonumber \\
    \mathbf{D} &= V^{vv}_{22}(\tau)
\end{align}
The result can be found in Eq. (\ref{eqn:inverse_schur}).

\section{Conventions and the Wiener-Hopf method}
\label{app:wiener-hopf}

Here, we first re-state the conventions used in dot products between functions operators and functions. Assume real functions of $t$, $f(t)$, $g(t)$, and a real operator $A(t-t')$, for $t,t' >0$. First, let us write down the inner product between the functions $f, g$: assuming that the functions are column vectors, we have
\begin{align}
    f^T \dotproduct g &\coloneqq \int_0^{\infty} f(t) \, g(t)\, dt \nonumber \\
    &= \int_{-\infty}^{\infty} f(t) \, g(t)\, dt \quad \text{if } f(t) \text{ or } g(t) \text{ is causal.} \nonumber \\
    &= \int_{-\infty}^{\infty} f(\Omega)^\dagger g(\Omega) \frac{d\Omega}{2\pi}
\end{align}
where $^T$ denotes the transpose operation such that $f^T$ is a row vector, $^{\dagger}$ denotes the complex conjugate operation, and $f(\Omega)$ is the Fourier transform of $f(t)$. Note that we used the Plancherel theorem in the last line. We use the following Fourier transform convention:
\begin{align}
    f(\Omega) &= \mathcal{F}\{f(t)\} = \int_{-\infty}^{\infty} dt\,f(t) e^{i \Omega t} \nonumber \\
    f(t) &= \mathcal{F}^{-1}\{f(t)\} = \int_{-\infty}^{\infty} \frac{d\Omega}{2\pi} f(\Omega) e^{-i \Omega t}
\end{align}
such that, for example,
\begin{align}
    &f(\Omega) = \frac{1}{b + i(a-\Omega)},\;a >0,\,b>0 \nonumber \\& \implies f(t) = e^{-(ia + b)t} H(t),
\end{align}
where $H(t)$ is the Heaviside step function. Then, we see that the spectrum of a causal (anti-causal) function has poles located on the lower (upper)-half of the complex plane.

The dot product between the operator $A$ and the function $f$ is given as
\begin{align}
    A \dotproduct f &= h(t) \coloneqq \int_0^{\infty} A(t'-t) \, f(t') \, dt', \, t>0 \nonumber \\
    f^T \dotproduct A &= (A^T \dotproduct f)^T = l(t) \coloneqq \int_0^{\infty} A(t-t') \, f(t') \, dt', \, t>0
\end{align}
where $h(t)$ and $l(t)$ are column and row vectors, respectively. Then, we see that the transpose operation corresponds to a time inversion for operators. Furthermore, if $f(t)$ is a causal function of $t$ (i.e. $f(t) = 0$ for $t<0$), we can change the integration to be between $-\infty$, $\infty$, and can write $h(\Omega) = A(-\Omega) f(\Omega) = A(\Omega)^\dagger f(\Omega)$, and $l(\Omega) = A(\Omega) f(\Omega)$. Note that $A(-\Omega) = A(\Omega)^\dagger$ since $A$ is a real operator.

Now, assume an operator $M$, and functions $h, g$, which are related to each other with
\begin{equation}
\label{supp_eqn:inner_prod}
   \int_0^\infty M(t-t')h(t') dt' = g(t), \quad t>0
\end{equation}
Eq. (\ref{supp_eqn:inner_prod}) implies that $h = M^{-1}  g$. In the main text, we want to find $h$ given $M$ and $g$, in order evaluate the characteristic polynomial (Eq. (\ref{eqn:determinant_full})). For this purpose, we use the Wiener-Hopf method (See Ref. \cite{Kisil2021} for a review). First, let us define $M(\Omega)$, $g(\Omega)$ as the spectrum of $M$ and $g$, respectively, and assume that they are rational functions of $\Omega$. Notice that Eq. (\ref{supp_eqn:inner_prod}) can be rewritten as
\begin{equation}
   \int_{-\infty}^\infty dt'\,M(t-t') h(t')_+ dt' = g(t), \quad t>0,
\end{equation}
where
\begin{equation}
f(t)_+ = \left[f(t)\right]_+ = \begin{cases}
       f(t) & \text{if } t > 0 \\
       0 & \text{otherwise.}
   \end{cases}
\end{equation}
which implies that
\begin{align}
\label{supp_eq:defining_s}  
& \left[\int_{-\infty}^\infty dt'\,\left[M(t-t')h(t')_+ - g(t)\right] \right]_+ = \left[s(t) \right]_+ = 0, \nonumber \\ 
   & s(t) \coloneqq \int_{-\infty}^\infty dt'\,\left[M(t-t')h(t')_+ - g(t)\right]
\end{align}
i.e. $s(t)$ is an anti-causal function of time. In the frequency domain, we have that
\begin{align}
    s(\Omega) &= \sum_{k} \frac{r(\Omega_k, \sigma_k)}{(\Omega-\Omega_k)^{\sigma_k}} = s(\Omega)_+ + s(\Omega)_-,
\end{align}
for any rational function $s(\Omega)$, where the sum is the partial fraction decomposition of the spectrum, and it is over the poles $\Omega_k$ of $s(\Omega)$, with multiplicity $\sigma_k$. $r(\Omega_k,\sigma_k)$ is the residue at $\Omega = \Omega_k$. Furthermore, $s(\Omega)_+$ is given by
\begin{align}
    s(\Omega)_+ &= \mathcal{F}\{s(t)_+\} \nonumber \\ &= \sum_{k} \frac{r(\Omega_k, \sigma_k)}{(\Omega-\Omega_k)^{\sigma_k}} \;\; \text{with} \;\; \text{Im}(\Omega_k) < 0
\end{align}
where $\text{Im}(\Omega_k)$ is the imaginary part of $\Omega_k$. From Eq. (\ref{supp_eq:defining_s}), we know that
\begin{align}
    s(\Omega) &= M(\Omega) h(\Omega)_+ - g(\Omega) \nonumber \\ \implies s(\Omega)_+ &= \left[ M(\Omega) h(\Omega)_+ - g(\Omega)\right]_+ = 0.
\end{align}
Since $M(\Omega)$ is a rational function, we can write it as $M(\Omega) = M(\Omega)_+ M(\Omega)_-$, where $M(\Omega)_+$ ($M(\Omega)_-$) contains the causal (anti-causal) zeros and poles of $M(\Omega)$. Furthermore, $M(\Omega)_- = M(\Omega)_+^{\dagger}$ since $M(-t) = M(t)^\dagger$ in the main text. We have
\begin{align}
    &\left[ M(\Omega)_+ M(\Omega)_- h(\Omega)_+ - g(\Omega)\right]_+ \nonumber \\
    &= \left[ M(\Omega)_+ h(\Omega)_+ - \frac{g(\Omega)}{M(\Omega)_-}\right]_+ = 0
\end{align}
where the second equality comes from the fact that multiplying the spectrum of an anti-causal function with a rational function with anti-causal roots and poles does not change the causality of the function. Therefore, we find 
\begin{align}
\label{eqn:wiener_hopf_2}
    h(\Omega)_+ &= \frac{1}{M(\Omega)_+} \left[\frac{g(\Omega)}{M(\Omega)_-}\right]_+, \nonumber \\ h(t) &= \mathcal{F}^{-1}\left\{ \frac{1}{M(\Omega)_+} \left[\frac{g(\Omega)}{M(\Omega)_-}\right]_+\right\},\;t>0.
\end{align}

\section{Existence of entanglement for all force noises in absence of sensing noise}
\label{app:no_sensing_noise}

Here, we want to obtain a condition for optomechanical entanglement in the presence of force noise only, i.e. in the absence of any sensing noise. % First, we recall from Sec. \ref{sec:entanglement_criterion} that optomechanical entanglement exists when $\mathbf{V}_{\text{pt}} + i \mathbf{K} < 0$, where $\mathbf{V}_{\text{pt}}$ is the covariance matrix of the system. 
Let us parametrize the spectrum of the force noise in the system as $\alpha_{\rm{F}} S_{\nf}$, where $\alpha_{\rm{F}}$ is a dimensionless constant scaling the overall strength of the noise; and $\alpha_{\rm{F}} = 0$ indicates that there is no force noise. 
%If the system is separable for a given $\alpha_{\rm{F}}$, it will remain separable for $\alpha_{\rm{F}} + \epsilon$, $\epsilon>0$. This is because the entanglement criterion becomes $\mathbf{V}_{\text{pt}} + \epsilon \mathbf{V}' + i \mathbf{K} < 0$, where $\mathbf{V}_{\text{pt}} + i \mathbf{K}$ is positive definite, and $\mathbf{V}'$ is positive semi-definite (since it is a covariance matrix). Therefore, the criterion is not satisfied, and optomechanical entanglement does not exist. 
If the system is entangled for $\alpha_{\rm{F}}$, it will also be entangled for $\alpha_{\rm{F}}-\epsilon$, since the entanglement criterion gets modified as $\mathbf{V}_{\text{pt}} + i \mathbf{K} - \epsilon \mathbf{V}' < 0$, where $\mathbf{V}_{\text{pt}} + i \mathbf{K}$ is negative definite, and $-\epsilon\mathbf{V}'$ is negative semi-definite (see Appendix \ref{app:uniqueness}). Then, the criterion will be satisfied for $\alpha_{\rm{F}}-\epsilon$. Therefore, we set $S_{\ns}(\Omega) = 0$ for the rest of the Appendix and characterize the force noise as $\alpha_{\rm{F}} S_{\nf}$, assuming that $\alpha_{\rm{F}} \gg 1$. Due to the discussion above, if the system is entangled in this limit, it will be entangled for all force noise spectra, regardless of the amplitude. 

In this limit, from Eq. (\ref{eqn:m_vac_defn}),
\begin{align}
 M(\Omega) &= \Omq^2|\chi(\Omega)|^2 \left( \omegam \alpha_{\rm{F}} S_{\nf}(\Omega) +2 \gammam \Omega \right) \nonumber \\ &= M_R(\Omega) + M_I(\Omega)  
\end{align}
where
\begin{align}
    M_R(\Omega) = \Omq^2|\chi(\Omega)|^2 \omegam \alpha_{\rm{F}} S_{\nf}(\Omega) = \Omq^2 \chiF(\Omega), \nonumber \\
    M_I(\Omega) = 2 \Omq^2|\chi(\Omega)|^2 \gammam \Omega = 2 \Omq^2 \Theta(\Omega) \gammam \Omega
\end{align}
Note that in the time domain, $M_R(t) = \Omq^2 \chiF(t)$, $M_I(t) = 2i \Omq^2 \gammam\dot \Theta(t)$. Since $M_R(\Omega)$ is a real and even function of $\Omega$, $M_R(t)$ is a real and symmetric function of time. Similarly, since $M_I(\Omega)$ is a real and odd function of $\Omega$, $M_I(t)$ is a purely imaginary and anti-symmetric function. Now, we want to express the inverse of the operator $M$ as a function of the operators $M_R$ and $M_I$, where $M_R$ is real and symmetric, and $M_I$ is imaginary and Hermitian (as $M_I(\tau) = -M_I(-\tau)$). In the limit of $\alpha_{\rm{F}} \gg 1$, we have
\begin{align}
     M^{-1} &= M_R^{-1} \dotproduct \left(1 + M_I \dotproduct M_R^{-1} \right)^{-1} \nonumber \\
     &\approx M_R^{-1} \dotproduct \left(1 - M_I \dotproduct M_R^{-1} \right)
     \nonumber \\ &= M_R^{-1} - M_R^{-1} \dotproduct M_I \dotproduct M_R^{-1}\label{eqn:approx_M-1}
\end{align}
where the approximation used that $M_R^{-1} \dotproduct M_I \propto 1/\alpha_{\rm{F}} \ll 1$.

With this approximation of $M^{-1}$, we evaluate Eq. (\ref{eqn:entanglement_indicator}). In order to calculate the dot products with $M_R^{-1}$ using the Wiener-Hopf method, we write $M_R^{-1}$ as a product of its causal and anti-causal parts. Note that we omit writing $\alpha_{\rm{F}}$ for the rest of the Appendix. First, we express $\chiF(\Omega)$ as a product of its causal and anti-causal parts:
\begin{align}
    \chiF(\Omega) &= \chiF(\Omega)_+\chiF(\Omega)_-, \nonumber \\ \chiF(\Omega)_+ &= \sqrt{\omegam} \chi(\Omega) S_{\nf}(\Omega)_+,
\end{align}
and $\chiF(\Omega)_- = \chiF(\Omega)_+^\dagger$.
Here, $S_{\nf} = S_{\nf}(\Omega)_+ S_{\nf}(\Omega)_-$, where $S_{\nf}(\Omega)_+$ contains the causal roots and zeros of $S_{\nf}(\Omega)$. This decomposition is possible for any power noise spectrum $S_{\nf}(\Omega)$ that is a rational function of $\Omega$. Then, we have: $M_R(\Omega)_\pm = \Omq\chiF(\Omega)_\pm$. 

We now want to calculate $\widetilde{\mathbf{V}}^{bv}_{\text{pt}} \dotproduct \mathbf{V}_M \dotproduct (\widetilde{\mathbf{V}}^{bv}_{\text{pt}})^T$ in Eq. (\ref{eqn:entanglement_indicator}). For this purpose, we can define $f \star g = f \dotproduct M^{-1} \dotproduct g$ for functions $f$ and $g$. Using this notation, $\widetilde{\mathbf{V}}^{bv}_{\text{pt}} \dotproduct \mathbf{V}_M \dotproduct (\widetilde{\mathbf{V}}^{bv}_{\text{pt}})^T$ is given by (we drop the superscripts $bv$)
\begin{align}
\label{supp_eqn:VbvMInvVvb_full_expression}
&\begin{bmatrix}
        V_{11} \star  V_{11}^T & - V_{11} \star  V_{21}^T \\[1mm] - V_{21} \star  V_{11}^T & V_{21} \star  V_{21}^T
    \end{bmatrix} + \begin{bmatrix} \widetilde{V}_{12} \star  \widetilde{V}_{12}^T
        & -\widetilde{V}_{12} \star  \widetilde{V}_{22}^T \\[1mm] -\widetilde{V}_{22} \star  \widetilde{V}_{12}^T & \widetilde{V}_{22} \star  \widetilde{V}_{22}^T
    \end{bmatrix} \nonumber \\  &- i\begin{bmatrix}
        V_{11} \star  \widetilde{V}_{12}^T - \widetilde{V}_{12} \star  V_{11}^T & \;\; \widetilde{V}_{12} \star  V_{21}^T - V_{11} \star  \widetilde{V}_{22}^T \\[1mm] \widetilde{V}_{22} \star  V_{11}^T - V_{21} \star  \widetilde{V}_{12}^T & \;\; V_{21} \star  \widetilde{V}_{22}^T - \widetilde{V}_{22} \star  V_{21}^T
    \end{bmatrix}
\end{align}

We first calculate Eq. (\ref{supp_eqn:VbvMInvVvb_full_expression}) by substituting $M^{-1}$ with $M_R^{-1}$ form Eq.~(\ref{eqn:approx_M-1}). We notice that
\begin{align}
    \widetilde{V}^{bv}_{12}(\Omega) &= \sqrt{\omegam} \Omq \chiF(\Omega)_+\chiF(\Omega)_-, \nonumber \\ 
    \widetilde{V}^{bv}_{22}(\Omega) &= -i\Omega \frac{\Omq}{\sqrt{\omegam}}  \chiF(\Omega)_+\chiF(\Omega)_-. 
\end{align}
We proceed with the Wiener-Hopf method to calculate $\widetilde{V}^{bv}_{12} \dotproduct M_R^{-1}$. In the frequency domain, we have
\begin{align}
(\widetilde{V}^{bv}_{12} \dotproduct M_R^{-1})(\Omega) &=  \frac{\sqrt{\omegam} \Omq}{\Omq^2} \frac{1}{\chiF(\Omega)_+} \left[\frac{\chiF(\Omega)_+\chiF(\Omega)_-}{\chiF(\Omega)_-}\right]_+ \nonumber \\ &= \frac{\sqrt{\omegam}}{\Omq}
\end{align}
therefore, in the time domain,
\begin{align}
\label{eqn:m_inv_1}
    \widetilde{V}^{bv}_{12} \dotproduct M_R^{-1} = 2\delta(t)\sqrt{\omegam}/\Omq
\end{align} 
for $t>0$. The factor of 2 arises from the causality of the delta function. Then, the inner product $\widetilde{V}^{bv}_{12} \dotproduct M_R^{-1} \dotproduct  (\widetilde{V}^{bv}_{12})^T$ is given by
\begin{align}
\widetilde{V}^{bv}_{12} \dotproduct M_R^{-1} \dotproduct  (\widetilde{V}^{bv}_{12})^T &= \frac{ 2\sqrt{\omegam}}{\Omq} \int_{0}^{\infty} \widetilde{V}^{bv}_{12}(t) \delta(t) dt \nonumber \\ &= \omegam \chiF(0) = (\widetilde{\mathbf{V}}^{bb})_{11}
\end{align}
Similarly, 
\begin{align}
\label{eqn:m_inv_2}
    \widetilde{V}^{vb}_{22} \dotproduct M_R^{-1} = \frac{2\dot \delta(t)}{\sqrt{\omegam}\Omq}
\end{align}
which implies that the second term in Eq. (\ref{supp_eqn:VbvMInvVvb_full_expression}) is given by 
\begin{align}
    &\begin{bmatrix} \;\;\,\widetilde{V}^{bv}_{12} \dotproduct M_R^{-1} \dotproduct  (\widetilde{V}^{bv}_{12})^T
        & -\widetilde{V}^{bv}_{12} \dotproduct M_R^{-1} \dotproduct  (\widetilde{V}^{bv}_{22})^T \\[1mm] -\widetilde{V}^{bv}_{22} \dotproduct M_R^{-1} \dotproduct  (\widetilde{V}^{bv}_{12})^T & \;\;\, \widetilde{V}^{bv}_{22} \dotproduct M_R^{-1} \dotproduct  (\widetilde{V}^{bv}_{22})^T
    \end{bmatrix}
    \nonumber \\ &= \begin{bmatrix}
        \omegam \chiF(0) & 0 \\
        0 & \frac{1}{\omegam} \ddot \chi_{\rm{F}}(0)
    \end{bmatrix} = \widetilde{\mathbf{V}}^{bb}
\end{align}
which cancels out with $\widetilde{\mathbf{V}}^{bb}$ in Eq. (\ref{eqn:entanglement_indicator}). Next, let us evaluate the third term in Eq. (\ref{supp_eqn:VbvMInvVvb_full_expression}). 
From Eq. (\ref{eqn:m_inv_1}) and Eq. (\ref{eqn:m_inv_2}), we have
\begin{align}
&\widetilde{V}^{bv}_{12} \dotproduct M_R^{-1} \dotproduct  (V^{bv}_{21})^T - \widetilde{V}^{bv}_{11} \dotproduct M_R^{-1} \dotproduct  (\widetilde{V}^{bv}_{22})^T \nonumber \\ &= \left(\frac{\sqrt{\omegam}}{\Omq} V^{bv}_{21}(0^+) + \frac{\dot V^{bv}_{11}(0^+)}{\sqrt{\omegam}\Omq} \right) = 2 \dot \chi(0^+) = 2
\end{align}
and the entries in the diagonals are zero, since $M_R$ is a real and symmetric operator.
Then, the third term in Eq. (\ref{supp_eqn:VbvMInvVvb_full_expression}) will be
\begin{align}
\begin{bmatrix}
        0 & -2i \\
        2i & 0
    \end{bmatrix}
\end{align}

Finally, we want to calculate the first term in Eq. (\ref{supp_eqn:VbvMInvVvb_full_expression}). However, one cannot directly take the dot products of $M_R^{-1}$ with $V^{bv}_{11}$ and $V^{bv}_{21}$: indeed if the force noise spectrum decays faster than $1/\Omega^2$, the expressions are not well-defined. However, (quantum) physically, we expect that any (bosonic) continuous variable force noise will remain finite; as we shall see at the end of this appendix. Then, we work in the limit of $M_R(t) = \epsilon + \Omq^2\chiF (t)$, $\epsilon \ll 1$. Similar to above, we need to write $M_R(t)$ as a product of a causal and anticausal function in the Fourier domain. For this purpose, notice that the spectrum of $M_R$ is---up to some constant---the spectrum of the force noise, $S_{\nf}(\Omega)$, multiplied by the absolute value squared of the mechanical susceptibility, $|\chi(\Omega)|^2$, due to the response of the harmonic oscillator to the external forces caused by the force noise. The force noise spectra can be written as a product of rational functions of $\Omega^2$ as it is a real and even function of $\Omega$. We assume that $S_{\nf}(\Omega)$ has $z$ zeros and $p$ poles, $k = p - z \geq 0$, $k$ even. First, notice that we can always write $S_{\nf}(\Omega)$ in the form of
\begin{align}
S_{\nf}(\Omega) \propto \frac{A(\Omega)_+A(\Omega)_-}{B(\Omega)_+B(\Omega)_-} = \frac{|A(\Omega)_+|^2}{|B(\Omega)_+|^2}
\end{align}
since $S_{\nf}(\Omega)$ is real. We set the coefficient multiplying $\Omega^2$ to the largest power in $B(\Omega)_+ B(\Omega)_-$ to unity, since any coefficient can be absorbed into $A(\Omega)_\pm$.
Furthermore, we define $D(\Omega)_+ = \Omega^2 + i\gammam \Omega - \omegam^2 =-1/\chi(\Omega)$, $D(\Omega)_- = D(\Omega)_+^\dagger$, which contain the poles of the mechanical susceptibility. Combining the new definitions, we write
\begin{align}
M_R(\Omega) &= \epsilon + \frac{1}{D(\Omega)_+D(\Omega)_-} \frac{A(\Omega)_+A(\Omega)_-}{B(\Omega)_+B(\Omega)_-} \nonumber \\ &\approx \frac{\epsilon \,\Omega^{4+p} + A(\Omega)_+A(\Omega)_-}{D(\Omega)_+D(\Omega)_-B(\Omega)_+B(\Omega)_-}
\end{align}
where the approximation holds in the limit of $\epsilon \ll 1$. Now, in order to find the roots of the numerator, notice that in the limit $\epsilon \rightarrow 0$, the numerator goes to $A(\Omega)_+A(\Omega)_-$. Then, we can write it as a product of $A(\Omega)_+A(\Omega)_-$ and the rest of the roots, which there are $4+k$ of them. To find the remaining roots, notice that they will be solutions to
\begin{align}
\label{eqn:force_noise_numerator_roots}
\epsilon \,\Omega^{4+p} + L \Omega^{z} = 0
\end{align}
where $L$ is a real number. since the roots will be at large $\Omega$ in the limit of $\epsilon \rightarrow 0$, and only the term $L\Omega^{z}$ in $A(\Omega)_+A(\Omega)_-$ will be of importance. The roots of Eq. (\ref{eqn:force_noise_numerator_roots}) are $\Omega_n = L \epsilon^{-1/(4+k)} \text{exp}((i2\pi n + i\pi)/(4+k))$, $n = 0,1,..., 3+k$. Then,
\begin{align}
M_R(\Omega) &\approx \frac{ A(\Omega)_+A(\Omega)_- \prod_{n=0}^{3+k} (\epsilon^{1/(4+k)} \Omega - Le^{\frac{i\pi}{4+k}+\frac{i2\pi n}{4+k}})}{D(\Omega)_+D(\Omega)_-B(\Omega)_+B(\Omega)_-} \nonumber \\
M_R(\Omega)_\pm &\approx \frac{A(\Omega)_\pm \prod_{n=0}^{1+k/2} (\epsilon^{1/(4+k)} \Omega \pm L e^{\frac{i\pi}{4+k}+\frac{i2\pi n}{4+k}})}{D(\Omega)_\pm B(\Omega)_\pm}
\end{align}
While using the Wiener-Hopf method, one needs to calculate $1/M_R(\Omega)_+$. We see that thanks to the terms with $\epsilon$, the number of poles and zeros of $1/M_R(\Omega)_+$, in the limit of $\epsilon \rightarrow 0$, are equal. Therefore, using Eq. (\ref{eqn:wiener_hopf_2}), the spectrum of $g(\Omega)_+$ is guaranteed to be bounded for bounded $h(\Omega)$. 

We now return to the evaluation of the first term in Eq.~(\ref{supp_eqn:VbvMInvVvb_full_expression}). In taking the dot products of $M_R^{-1}$ with $V^{bv}_{11}$ and $V^{bv}_{21}$ one encounters expressions such as
\begin{align}
\label{supp_eqn:M_inner_prod_1}
 V^{bv}_{11} \dotproduct M_R^{-1} \dotproduct  V^{vb}_{11} = \int \frac{d \Omega}{2 \pi} \frac{V^{vb}_{11}(\Omega)^\dagger}{M_R(\Omega)_+} \left[ \frac{V^{bv}_{11}(\Omega)}{M_R(\Omega)_-} \right]_+
\end{align}
Since the number of poles and zeros of $M_R(\Omega)_+$ in the limit of $\epsilon \rightarrow 0$ are equal from the discussion, and $V^{bv}_{11}(\Omega)$ has two poles, we can write this as the dot product of two bounded, causal functions of time. Or, in the frequency domain, as
\begin{align}
\label{supp_eqn:M_inner_prod_2}
 \int \frac{d \Omega}{2 \pi} \left[ \frac{V^{bv}_{11}(\Omega)^\dagger}{M_R(\Omega)_+} \right]_- \left[ \frac{V^{bv}_{11}(\Omega)}{M_R(\Omega)_-} \right]_+
\end{align}
Then, we only need the residues of $1/M_R(\Omega)_-$ at the poles of the mechanical susceptibility to compute the dot products in the limit of $\epsilon \rightarrow 0$. In the end, the first term in Eq. (\ref{supp_eqn:VbvMInvVvb_full_expression}) is
\begin{align}
        \begin{bmatrix}
        \;\;\, V^{bv}_{11} \dotproduct M_R^{-1} \dotproduct (V^{bv}_{11})^T & -V^{bv}_{11} \dotproduct M_R^{-1} \dotproduct (V^{bv}_{21})^T \\[1mm] -V^{bv}_{21} \dotproduct M_R^{-1} \dotproduct (V^{bv}_{11})^T & \;\;\, V^{bv}_{21} \dotproduct M_R^{-1} \dotproduct (V^{bv}_{21})^T
    \end{bmatrix}
\end{align}

Now that we have finished discussing the terms with $M_R^{-1}$, we want to calculate $\widetilde{\mathbf{V}}^{bv}_{\text{pt}} \dotproduct \mathbf{V}_M \dotproduct\widetilde{\mathbf{V}}^{vb}_{\text{pt}}$ (Eq. (\ref{supp_eqn:VbvMInvVvb_full_expression})) again, substituting $M^{-1}$ with $M_R^{-1} \dotproduct M_I \dotproduct M_R^{-1}$; cf. Eq.~(\ref{eqn:approx_M-1}). First, note that after this substitution, the first term in Eq. (\ref{supp_eqn:VbvMInvVvb_full_expression}) will be proportional to $1/\alpha_{\rm{F}}^2 \ll 1$, therefore we can ignore this term. From Eqs. (\ref{eqn:m_inv_1}), (\ref{eqn:m_inv_2}), the second term will be given by 
\begin{align}
    \frac{1}{\Omq^2} \begin{bmatrix}
        \omegam M_I(0) & \dot M_I(0) \\ -\dot M_I(0) & \frac{-1}{\omegam} \ddot M_I(0)
    \end{bmatrix} = \begin{bmatrix}
        0 & -i \\ i & 0
    \end{bmatrix}
\end{align}
since $M_I(0) = \ddot M_I(0) = 0$. In order to calculate the third term in Eq. (\ref{supp_eqn:VbvMInvVvb_full_expression}), we notice that
\begin{align}
    & \widetilde{V}^{bv}_{12} \dotproduct M_R^{-1} \dotproduct M_I = \delta(t) \dotproduct M_I \frac{\sqrt{\omegam}}{\Omq} = M_I(t) \frac{\sqrt{\omegam}}{\Omq}  \nonumber \\
    & \widetilde{V}^{bv}_{22} \dotproduct  M_R^{-1} \dotproduct M_I = \dot \delta(t) \dotproduct M_I \frac{1}{\sqrt{\omegam}\Omq} = \frac{\dot M_I(t)}{\sqrt{\omegam}\Omq}
\end{align}
for $t>0$. Then, the third term in Eq. (\ref{supp_eqn:VbvMInvVvb_full_expression}) will involve inner products with $M_I(t)$ and $\dot M_I(t)$. Since these products will have support only on the half real line (i.e. the integration limits will be from $0$ to $+\infty$), we can replace $M_I(t)$ and $\dot M_I(t)$ with their causal counterparts. Notice that
\begin{align}
    M_I(t) \frac{\sqrt{\omegam}}{\Omq} = -i\Omq \sqrt{\omegam} \chi(t) = -iV^{bv}_{11}(t)
\end{align}
for $t>0$. Similarly, $\dot M_I(t)/\sqrt{\omegam}\Omq = -iV^{bv}_{21}(t)$ for $t>0$. Therefore, the third term in Eq. (\ref{supp_eqn:VbvMInvVvb_full_expression}) can be written as
\begin{align}
    2\begin{bmatrix}
          V^{bv}_{11} \dotproduct M_R^{-1} \dotproduct (V^{bv}_{11})^T & -  V^{bv}_{11} \dotproduct M_R^{-1} \dotproduct (V^{bv}_{21})^T \\[1mm]  -V^{bv}_{21} \dotproduct M_R^{-1} \dotproduct (V^{bv}_{11})^T & \;\; V^{bv}_{21} \dotproduct M_R^{-1} \dotproduct (V^{bv}_{21})^T
    \end{bmatrix}
\end{align}
Therefore, summing all of the matrices above, the condition to be entangled in the absence of sensing noise is given by
\begin{align}
\label{eq:vacuum_inner_upper_bound}
\text{det} \Biggl\{ &\begin{bmatrix}
         V^{bv}_{11} \dotproduct M_R^{-1} \dotproduct (V^{bv}_{11})^T &  -V^{bv}_{11} \dotproduct M_R^{-1} \dotproduct (V^{bv}_{21})^T \\[1mm]   -V^{bv}_{21} \dotproduct M_R^{-1} \dotproduct (V^{bv}_{11})^T & V^{bv}_{21} \dotproduct M_R^{-1} \dotproduct (V^{bv}_{21})^T
    \end{bmatrix} \nonumber \\
    &+ \begin{bmatrix}
        0 & 2i \\
        -2i & 0
    \end{bmatrix} \Biggr\}<0
\end{align}
using the Wiener-Hopf method and the expressions for $V^{bv}_{11}$, $V^{bv}_{21}$, and $M_R$, where we calculate the following elegant expression:
\begin{align}
\label{eq:vacuum_inner}
&(V^{bv}_{11} \dotproduct M_R^{-1} \dotproduct  (V^{bv}_{11})^T)(V^{bv}_{21} \dotproduct M_R^{-1} \dotproduct  (V^{bv}_{12})^T) \nonumber \\ &- (V^{bv}_{11} \dotproduct M_R^{-1} \dotproduct  (V^{bv}_{21})^T)^2 = \frac{4 \gammam^2}{|S_{\nf-}(\Omega_*)|^4}
\end{align}
where $\Omega_* = -i\gammam/2 \pm \sqrt{\omegam^2-\gammam^2/4}$ is one of the poles of the mechanical susceptibility. From Eqs. (\ref{eq:vacuum_inner_upper_bound}) and (\ref{eq:vacuum_inner}), the condition for optomechanical entanglement reduces to
\begin{align}
\label{supp_eqn:no_sensing_entanglement_criterion}
    |S_{\nf-}(\Omega_*)|^2 > \gammam \iff \text{entanglement}.
\end{align}
There exists a lower limit to $|S_{\nf-}(\Omega_*)|$ from the fluctuation-dissipation theorem \cite{callen_1951, Kubo_1966}, since $S_{\nf}(\Omega)$ includes the thermal-fluctuations that give rise to velocity damping on the mechanical oscillator, with a damping rate $\gammam$ defined in the main text. Then, assuming that there is no additional force noise in the system, $S_{\nf}(\Omega)$ is connected to the mechanical susceptibility with
\begin{align}
S_{\nf}(\Omega)|\chi(\Omega)|^2 = \frac{2 \gammam \Omega}{\omegam} |\chi(\Omega)|^2 \coth{\left(\frac{\hbar \Omega}{2k_BT}\right)}
\end{align}
where $T$ is the temperature of the bath, and $k_B$ is the Boltzmann constant. Eliminating the mechanical susceptibility from each side, and writing $S_{\nf}$ as $S_{\nf}(\Omega) = S_{\nf}(\Omega)_+S_{\nf}(\Omega)_-$, the minimum value of $|S_{\nf-}(\Omega_*)|^2$ is given by
\begin{align}
|S_{\nf-}(\Omega_*)|^2 \geq 2 \gammam
\end{align}
We therefore see that all physical force noise spectra ensure optomechanical entanglement, from Eq. (\ref{supp_eqn:no_sensing_entanglement_criterion}).

\section{Uniqueness of the entangling-disentangling transition}
\label{app:uniqueness}

Here, we assume a force and sensing noise spectrum $\alpha_{\rm{F}} S_{\nf}(\Omega)$, $\beta_{\rm{S}} S_{\ns}(\Omega)$, respectively, where $\alpha_{\rm{F}}$ and $\beta_{\rm{S}}$ are real, positive constants. We want to show that given $\alpha_{\rm{F}}$, the value of $\beta_{\rm{S}}$ for which the entangling-disentangling transition occurs is unique. In other words, if the system is not entangled for a pair $(\alpha_{\rm{F}}, \beta_{\rm{S}})$, entanglement cannot be achieved by increasing $\beta_{\rm{S}}$. Note that when we increase $\beta_{\rm{S}}$, we introduce more sensing noise to the system, however, we do not change the functional form of the spectrum (i.e. $S_{\ns}(\Omega)$). 

We write the covariance matrix as $\mathbf{V}$, and add some additional sensing noise to the system, modifying the covariance matrix in the form of $\mathbf{V} \rightarrow \mathbf{V} + r \mathbf{V}'$, where $r \in \mathbb{R}$, and $\mathbf{V}'$ is positive-semi-definite as it is a covariance matrix. $\mathbf{V}'$ will be non-zero only in the sector of the covariance matrix containing the variance of $\hat v_2$ due to the structure of the equations of motion, given in Eq. (\ref{eqn:eqns_time_domain}). This modification corresponds to modifying $\beta_{\rm{S}}$, such that $\beta_{\rm{S}} \rightarrow \beta_{\rm{S}} + r$.

We recall from Sec. \ref{sec:entanglement_criterion} that optomechanical entanglement exists when $\mathbf{V}_{\text{pt}} + i \mathbf{K} < 0$, where $\mathbf{V}_{\text{pt}}$ is the covariance matrix of the system. If the system is separable for a given $\beta_{\rm{S}}$, it will remain separable for $\beta_{\rm{S}} + r$, $r>0$. This is because the entanglement criterion becomes $\mathbf{V}_{\text{pt}} + r \mathbf{V}' + i \mathbf{K} < 0$, where $\mathbf{V}_{\text{pt}} + i \mathbf{K}$ is positive definite, and $\mathbf{V}'$ is positive semi-definite (since it is a covariance matrix). Therefore, the criterion is not satisfied, and optomechanical entanglement does not exist. Conversely, if optomechanical entanglement exists for $\beta_{\rm{S}}$, it will also exist for $\beta_{\rm{S}}-r$, $r>0$, since the criterion gets modified as $\mathbf{V}_{\text{pt}} + i \mathbf{K} - r \mathbf{V}' < 0$, where $\mathbf{V}_{\text{pt}} + i \mathbf{K}$ is negative definite, and $-\mathbf{V}'$ is negative semi-definite. Then, the criterion will be satisfied for $\beta_{\rm{S}}-r$. 

Therefore, if we introduce some sensing noise to an initially entangled system, there exists a critical amount of additional sensing noise for which an entangling-disentangling transition occurs, and the system becomes separable. Beyond this point, the system cannot get entangled again as we continue introducing more sensing noise. Conversely, before the entangling-disentangling transition (when the system is entangled), we cannot induce separability by removing sensing noise from the system. These statements prove our initial claim that the entangling-disentangling transition is unique with respect to the amount of sensing noise present.

\section{Another approach to proving the universality of entanglement}
\label{app:yanbei_proof}

Let us provide another approach to proving the universality of entanglement when the input light field is the vacuum state, as was shown in Sec. \ref{sec:universal_entanglement}. The entanglement criterion that we use in the main text is
\begin{align}
    \text{det}\left( \mathbf{V}_{\text{pt}} + i \mathbf{K} \right) < 0 \iff \text{entanglement}.
\end{align}
We can apply a transformation $\mathbf{T}$ to this expression that is not necessarily symplectic, such that the entanglement criterion is modified as
\begin{align}
    \text{det}\left( \mathbf{T} \mathbf{V}_{\text{pt}} \mathbf{T}^\dagger + i \mathbf{T} \mathbf{K} \mathbf{T}^\dagger \right) < 0 \iff \text{entanglement}.
\end{align}
since $\mathbf{T}$ is not symplectic, we cannot use the transformed expression to quantify the amount of optomechanical entanglement, however, we can still inquire the existence of entanglement. We choose $\mathbf{T}$ to be
\begin{align}
    &\quad \mathbf{T} = \begin{bmatrix}
        \mathbf{T}^{bb} & \mathbf{T}^{bv} \\
        \mathbf{T}^{vb} & \mathbf{T}^{vv}
    \end{bmatrix}, \nonumber \\
    \mathbf{T}^{bb} &= \begin{bmatrix}
        1 & 0 \\
        0 & 1
    \end{bmatrix}, \;\;\, \mathbf{T}^{bv} = \sqrt{\omegam} \Omq \begin{bmatrix}
        -\chi(t) & 0 \\[1mm]
        \frac{1}{\omegam} \dot \chi(t) & 0
    \end{bmatrix}, \nonumber \\
    \mathbf{T}^{vb} &= \mathbf{0}, \quad \quad \quad \mathbf{T}^{vv} = \begin{bmatrix}
        \delta(\tau) & 0 \\
        - \Omq^2 \chi(\tau) & \delta(\tau)
    \end{bmatrix}.
\end{align}
Then, $\mathbf{T} \mathbf{V}_{\text{pt}} \mathbf{T}^\dagger$ is calculated as
\begin{align}
    \begin{bmatrix}
        \omegam \chiF(0) & 0 & 0 & \sqrt{\omegam} \Omq \chi_{\rm{F}}(t') \\[1mm]
        0 & \frac{-1}{\omegam} \ddot \chiF(0) & 0 & -\frac{\Omq}{\sqrt{\omegam}} \dot \chi_{\rm{F}}(t') \\
        0 & 0 & \delta(\tau) & 0 \\
        \sqrt{\omegam} \Omq \chi_{\rm{F}}(t) & -\frac{\Omq}{\sqrt{\omegam}} \dot \chi_{\rm{F}}(t) & 0 & \delta(\tau) + \Omq^2 \chi_{\text{EN}}(\tau)
    \end{bmatrix}
\end{align}
where $t, t' >0$ are the column and the row indices, respectively, and $\tau = t - t'$. This matrix can be viewed as the partially transposed covariance matrix of the quadratures $\hat c_i, \hat \omega_i(t)$, $i = 1, 2$, $t <0$. These obey the following equations of motion:
\begin{subequations}
\label{app_eqn:transformed_eqns_motion}
\begin{align}
    \hat{\omega}_1(t) &= \hat{u}_1(t) \,, \\
\hat{\omega}_2(t) &= \hat{u}_2(t) + \frac{\Omq}{\sqrt{\omegam}} (\hat{c}_1(t)+\hat{n}_{\rm{S}}(t)) \,, \\
 \dot{\hat{c}}_2(t) &= -\gammam \hat{c}_2(t) -\omegam \hat{c}_1(t) +\hat{n}_{\rm{F}}(t), \\
\dot{\hat{c}}_1(t) &= \omegam \hat{c}_2(t) \,
\end{align}
\end{subequations}
Comparing to the original equations of motion, Eq. (\ref{eqn:eqns_time_domain}), we notice that here, the mechanical mode $\hat c_1, \hat c_2$ is driven only by the force noise, $\hat{n}_{\rm{F}}(t)$. 

We compute $\mathbf{T} \mathbf{K} \mathbf{T}^\dagger$ as
\begin{align}
    \begin{bmatrix}
        0 & 1 & 0 & -\sqrt{\omegam} \Omq \chi(t') \\[1mm]
        -1 & 0 & 0 & \frac{\Omq}{\sqrt{\omegam}} \dot \chi(t')  \\[1mm]
        0 & 0 & 0 & \delta(\tau) \\[1mm]
        \sqrt{\omegam} \Omq \chi(t) & -\frac{\Omq}{\sqrt{\omegam}} \dot \chi(t) & -\delta(\tau) & \Omq^2 (\chi(-\tau) - \chi(\tau))
    \end{bmatrix}
\end{align}
where, again, $t, t' >0$, and $\tau = t - t'$. Therefore, we can write $\mathbf{T} \mathbf{V}_{\text{pt}} \mathbf{T}^\dagger + i \mathbf{T} \mathbf{K} \mathbf{T}^\dagger$ as (using the same row and column indices $t$ and $t'$, as well as $\tau = t - t'$),
\begin{widetext}
\begin{align}
\label{app_eqn:yanbei_cov_1}
\begin{bmatrix}
    \langle \hat c^2_1(0) \rangle & i & 0 & \langle \hat c_1(0) \hat \omega_2(t') \rangle_s -i \sqrt{\omegam}\Omq \chi(t')  \\[1mm]
        -i & \langle \hat c^2_2(0) \rangle & 0 & -  \langle \hat c_2(0) \hat \omega_2(t') \rangle_s + i \frac{\Omq}{\sqrt{\omegam}} \dot \chi(t')  \\
        0 & 0 & \delta(\tau) & i \delta(\tau) \\
         \langle \hat c_1(0) \hat \omega_2(t) \rangle_s + i \sqrt{\omegam}\Omq \chi(t) & -  \langle \hat c_2(0) \hat \omega_2(t) \rangle_s - i \frac{\Omq}{\sqrt{\omegam}} \dot \chi(t) & -i \delta(\tau) & \delta(\tau) + M(\tau)
\end{bmatrix}
\end{align}
\end{widetext}
where we define the symmetrized expectation value $\langle \hat a \hat b \rangle_s \coloneqq \langle \hat a \hat b + \hat b \hat a \rangle/2$, and $M(\tau) \coloneqq \langle \hat c_1(t) \hat c_1(t') \rangle \Omq^2/\omegam + \langle \hat n_{\rm{S}}(t) \hat n_{\rm{S}}(t') \rangle$, which is also equivalent to the expression that we had for $M(\tau)$ in the main text, given in Eq. (\ref{eqn:m_vac_defn}). %Furthermore, we used the fact that $\langle \hat c_i(0) \hat v_2(t) \rangle_s = \langle \hat c_i(0) \hat c_1(t) \rangle_s \, \Omq/\sqrt{\omegam}$ for $t<0$, $i = 1, 2$, as the quadratures $\hat c_1, \hat c_2$ are not driven by the sensing noise. Similarly, $\langle \hat v_2(t) \hat v_2(t') \rangle = \langle \hat c_1(t) \hat c_1(t') \rangle \Omq^2/\omegam + \langle \hat n_{\rm{S}}(t) \hat n_{\rm{S}}(t') \rangle$, $t, \, t' <0$.

We can simplify this expression by relating some of its terms to the commutation relations between the quadratures $\hat c_j(0)$ and $\hat \omega_j(t)$, $j = 1, 2$. First, we realize that
\begin{align}
\label{app_eqn:causality_comm}
    [\hat b_j(0), \hat v_2(t)] = 0, \;\; j = 1, 2, \;\; t<0.
\end{align}
due to causality. Furthermore, we can write
\begin{align}
\label{app_eqn:b_c}
    \hat b_1(t) &= \hat c_1(t) + \sqrt{\omegam} \Omq \int_{-\infty}^{t} \chi(-t') \hat u_1(t') \, dt' \nonumber \\
    \hat b_2(t) &= \hat c_2(t) + \frac{\Omq}{\sqrt{\omegam}}\int_{-\infty}^{t} \dot \chi(-t') \hat u_1(t') \, dt'
\end{align}
Therefore, using Eqs. (\ref{app_eqn:causality_comm}), (\ref{app_eqn:b_c}), as well as the original equations of motion in Eq. (\ref{eqn:eqns_time_domain}), we find
\begin{subequations}
\begin{align}
     [\hat b_j(0), \hat v_2(t)] &= [\hat c_j(0) + \Omq \omegam^{\frac{3}{2}-j} \int_{-\infty}^{\infty} \chi(-t') \hat u_1(t') dt', \nonumber \\ & \quad \hat \omega_2(t) + \Omq^2 \int_{-\infty}^t \chi(-t') \hat u_1(t') dt']
\end{align}
\end{subequations}
Since $\hat c_j$ does not contain any vacuum fluctuations, we can simplify this expression to obtain for $[\hat b_j(0), \hat v_2(t)]$
\begin{align}
    &[\hat c_j(0) ,\hat \omega_2(t)] + \Omq \omegam^{\frac{3}{2}-j} \int_{-\infty}^{\infty} dt' \chi(-t') [  \hat u_1(t'), \hat \omega_2(t)] \nonumber \\
    &= [\hat c_j(0) ,\hat \omega_2(t)] + \Omq \omegam^{\frac{3}{2}-j} \int_{-\infty}^{\infty} dt' \chi(-t') 2i \delta(t-t') \nonumber \\
    &= [\hat c_j(0) ,\hat \omega_2(t)] + 2i\Omq \omegam^{\frac{3}{2}-j}  \chi(-t) = 0
\end{align}
Therefore, we have
\begin{subequations}
\begin{align}
     [\hat c_1(0), \hat \omega_2(t)] &= - 2i \sqrt{\omegam} \Omq \chi(-t), \quad t<0 \\
     [\hat c_2(0), \hat \omega_2(t)] &= - 2i \frac{\Omq}{\sqrt{\omegam}} \dot \chi(-t), \quad t<0
\end{align}
\end{subequations}
We can then re-express the symmetrized expectation values in the form of
\begin{align}
    \langle \hat c_1(0) \hat \omega_2(t) \rangle_s &= \frac{\langle \hat c_1(0) \hat \omega_2(t) + \hat \omega_2(t) \hat c_1(0) \rangle}{2} \nonumber \\
    &= \langle \hat c_1(0) \hat \omega_2(t) \rangle + i \sqrt{\omegam} \Omq \chi(-t)
\end{align}
We can write a similar expression for $\langle \hat c_2(0) \hat \omega_2(t) \rangle_s$. Plugging these equations into Eq. (\ref{app_eqn:yanbei_cov_1}), we obtain
\begin{align}
\begin{bmatrix}
    \langle \hat c^2_1(0) \rangle & i & 0 & \langle \hat c_1(0) \hat \omega_2(t') \rangle\\[1mm]
        -i & \langle \hat c^2_2(0) \rangle & 0 & -\langle \hat c_2(0) \hat \omega_2(t') \rangle \\
        0 & 0 & \delta(\tau) & i \delta(\tau) \\
        \langle \hat \omega_2(t) \hat c_1(0) \rangle & - \langle \hat \omega_2(t) \hat c_2(0) \rangle & -i \delta(\tau) & \delta(\tau) + M(\tau)
\end{bmatrix}.
\end{align}
To inquire about optomechanical entanglement, we need to assess whether the smallest eigenvalue of this matrix is negative. Using the Schur complement of the top left block, we can rewrite the entanglement criterion as
\begin{align}
\label{app_eqn:yanbei_test}
    \lambda_{1,1} \lambda_{2,2} - \lambda_{1,2}^2 <1 \iff \text{entanglement},
\end{align}
where
\begin{align}
    \lambda_{i,j} &= \langle \hat c_i(0) \hat c_j(0) \rangle \nonumber \\
    &\quad- \int dt dt' \langle \hat c_i(0) \hat \omega_2(t') \rangle M^{-1}(\tau) \langle \hat \omega_2(t) \hat c_j(0) \rangle
\end{align}
with $\tau = t-t'$, $i, j = 1, 2$. The physical picture here contains a mechanical oscillator driven solely by a force noise $\hat n _{\rm{F}}$, sensed by a field that contains a sensing noise $\hat n _{\rm{S}}$. Then, optomechanical entanglement is observed only if the conditional quantum state of the mechanical oscillator, given its initial state and the output light field during $t<0$, violates the Heisenberg uncertainty principle. Accordingly, Eq. (\ref{app_eqn:yanbei_test}) contains the second-order moments of the conditional state, and tests if the uncertainty principle is held. Mathematically, this statement is exactly equivalent to Eq. (\ref{eqn:entanglement_indicator}) in the main text, and the interaction strength $\Omq$ cancels out of Eq. (\ref{app_eqn:yanbei_test}), as we had observed, giving rise to the universality of optomechanical entanglement.

\section{Universal entanglement in the presence of frequency-independent squeezing}
\label{app:squeezing_or_rotations}

We consider two classes of symplectic transformations on the input light field, (i) phase shifts, and (ii) squeezing the amplitude and phase quadratures with a constant squeezing factor. We show that both of these transformations result in the same determinant condition for optomechanical entanglement computed for the case when the input light field consists of vacuum fluctuations, given in Eq. (\ref{eqn:entanglement_indicator}). First, let us consider phase shift operations. Such a transformation can be represented by
\begin{align}
\label{supp_eqn:only_rotations}
    \begin{bmatrix}
        \hat u_1(\Omega) \\ \hat u_2(\Omega)
    \end{bmatrix} \rightarrow e^{i \phi(\Omega)} \begin{bmatrix}
\cos{\theta(\Omega)} & -\sin{\theta(\Omega)} \\ \sin{\theta(\Omega)} & \;\;\,\,\cos{\theta(\Omega)}
    \end{bmatrix} \cdot \begin{bmatrix}
        \hat u_1(\Omega) \\ \hat u_2(\Omega)
    \end{bmatrix}
\end{align}
where $\theta(\Omega)$ is a real function of $\Omega$, and the phase factor $e^{i \phi(\Omega)}$ ensures the causality of the transformation. We compute the light field sector of the covariance matrix in the frequency domain as
\begin{align}
    V^{vv}_{11}(\Omega) &= \cos^2{\theta(\Omega)} + \sin^2{\theta(\Omega)} = 1, \nonumber \\ V^{vv}_{12}(\Omega) &= \cos^2{\theta(\Omega)} \, \Omq^2 \chi(-\Omega) - \cos{\theta(\Omega)}\sin{\theta(\Omega)} \nonumber \\ &+ \cos{\theta(\Omega)}\sin{\theta(\Omega)} + \sin^2{\theta(\Omega)}\, \Omq^2 \chi(-\Omega) \nonumber \\ &= \Omq^2 \chi(-\Omega), \nonumber \\
    V^{vv}_{22}(\Omega) &= |\cos{(\theta(\Omega))} \, \Omq^2 \chi(\Omega) - \sin{(\theta(\Omega))}|^2 \nonumber \\ &+ |\sin{(\theta(\Omega))} \, \Omq^2 \chi(\Omega) + \cos{(\theta(\Omega))}|^2 + \Omq^2  \chi_\text{EN}(\Omega) \nonumber \\ &= 1 + \Omq^2 \chi_\text{EN}(\Omega) +  \Omq^4 |\chi(\Omega)|^2
\end{align}
Then, we see that the light-field sector of the covariance matrix is unchanged with respect to the case where the input light field is the vacuum state. Similarly, we find that all of the elements of the covariance matrix are unchanged. Hence, the condition for the force and the sensing noise spectra for which the entangling-disentangling transition occurs is unaltered.

Now, we consider squeezing the quadratures of the light field with a constant squeezing factor. Such a transformation looks like the following
\begin{align}
    \begin{bmatrix}
        \hat u_1(\Omega) \\ \hat u_2(\Omega)
    \end{bmatrix} \rightarrow \begin{bmatrix}
        e^r & 0 \\ 0 & e^{-r}
    \end{bmatrix} \cdot \begin{bmatrix}
        \hat u_1(\Omega) \\ \hat u_2(\Omega)
    \end{bmatrix}
\end{align}
where $r$ is a real number. In order to utilize the formalism of the main text, we can perform a symplectic transformation that simply rescales $\hat v_1(t)$ and $\hat v_2(t)$ as $\hat v_1(t) \rightarrow e^{-r}\hat v_1(t)$, $\hat v_2(t) \rightarrow e^{r}\hat v_2(t)$. Then, the light field sector of the covariance matrix is computed after the symplectic transformation as
\begin{align}
\label{supp_eqn:cov_vv_const_squeezing}
    & V^{vv}_{11}(\Omega) = 1, \quad V^{vv}_{12}(\Omega) = e^{2r} \Omq^2 \chi(-\Omega), \nonumber \\ &
    V^{vv}_{22}(\Omega) = 1 + e^{2r}\Omq^2 \chi_\text{EN}(\Omega) +  e^{4r}\Omq^4 |\chi(\Omega)|^2
\end{align}
The covariances of the quadratures of the mechanical oscillator are 
\begin{align}
\label{supp_eqn:cov_bb_const_squeezing}
    V^{bb}_{11}(\Omega) &= \omegam\chiF(\Omega) + \omegam\Omq^2 |\chi(\Omega)|^2 e^r, \nonumber \\ V^{bb}_{22}(\Omega) &= \frac{\Omega^2}{\omegam^2} V^{bb}_{11}(\Omega),
\end{align}
and $V^{bb}_{12}(0) = V^{bb}_{21}(0) = 0$ due to symmetry.
Lastly, the correlations between the quadratures of the mechanical oscillator and the light field are computed as
\begin{align}
\label{supp_eqn:cov_bv_const_squeezing}
    V^{bv}_{11}(\Omega&) = \sqrt{\omegam} e^r \Omq \chi(\Omega), \nonumber \\
    V^{bv}_{12}(\Omega) &= \sqrt{\omegam} e^r \Omq \,\chiF(\Omega) + \sqrt{\omegam} e^{3r} \Omq^3 |\chi(\Omega)|^2,
\end{align}
and $V^{bv}_{21}(\Omega) = -\frac{i\Omega}{\omegam} V^{bv}_{11}(\Omega)$, $V^{bv}_{22}(\Omega) = -\frac{i\Omega}{\omegam} V^{bv}_{12}(\Omega)$. From Eqs. (\ref{supp_eqn:cov_vv_const_squeezing}-\ref{supp_eqn:cov_bv_const_squeezing}), we notice that the effect of the frequency independent squeezing is such that the interaction strength $\Omq$ is rescaled to $e^r\Omq$. Since we have proved the universality of the entangling-disentangling transition with respect to the interaction strength (see Section \ref{sec:universal_entanglement}), rescaling it will not affect the transition. Therefore, we conclude that the condition for the force and the sensing noise spectra for which the entangling-disentangling transition occurs is unaltered if the input light field is squeezed by a constant squeezing factor.

\section{General symplectic transformations on the light field}
\label{app:general_symplectic_transformations}

We want to write down an expression for a general, causal symplectic transformation applied to the light field. Such a transformation can be written down as a 2-by-2 matrix of operators $\mathbf{S}$, with
\begin{align}
\left[\begin{array}{c}
     \hat{u}_1(t) \\
     \hat{u}_2(t)
\end{array}\right] \rightarrow \mathbf{S} \left[\begin{array}{c}
     \hat{u}_1(t) \\
     \hat{u}_2(t)
\end{array}\right], \quad \mathbf{S} = \left[\begin{array}{cc}
     S_{11}(\tau) \; & S_{12}(\tau) \\
     S_{21}(\tau) \; & S_{22}(\tau)
\end{array}\right]
\end{align}
where $S_{ij}(t)$ are causal functions of time, $i, j = 0, 1$. In order to be a valid symplectic transformation, $\mathbf{S}$ must satisfy \cite{cv_qi_adesso}
\begin{equation}
    \mathbf{S}^T \mathbf{K}\mathbf{S} = \mathbf{K}, \quad  \mathbf{K} = \left[\begin{array}{cc}
    0 & \, \delta(\tau) \\
    - \delta(\tau) \, & 0
\end{array}\right]
\end{equation}
where $\mathbf{K}$ is the commutator matrix. Writing $\mathbf{S}^T \mathbf{K}\mathbf{S} = \mathbf{K}$ explicitly,
\begin{widetext}
\begin{align}
     \int_{0}^{\infty} dt' \left[\begin{array}{cc}
     S_{11}(\tau_1) S_{21}(\tau_2) - S_{21}(\tau_1) S_{11}(\tau_2)  \; & \; S_{11}(\tau_1) S_{22}(\tau_2) - S_{21}(\tau_1) S_{12}(\tau_2) \\ S_{12}(\tau_1) S_{21}(\tau_2) - S_{22}(\tau_1) S_{11}(\tau_2) \; & \; S_{12}(\tau_1) S_{22}(\tau_2) - S_{22}(\tau_1) S_{12}(\tau_2)
\end{array}\right] = \left[\begin{array}{cc}
    0 & \, \delta(t-t'') \\
    - \delta(t-t'') \, & 0
\end{array}\right],
\end{align}
\end{widetext}
with $\tau_1 := t'-t, \; \tau_2 := t''-t'$. We can extend the lower limit of this integral to $-\infty$ due to the causality of $S_{ij}(t)$, $i, j = 1,2$. Thus, we can express these constraints in the frequency domain, by defining $S_{ij}(\Omega) = \int_{-\infty}^{\infty} d\tau \, S_{ij}(\tau) \, e^{i\Omega \tau}$, $i, j = 1,2$. We have:
\begin{subequations}
\begin{align}
    &S_{11}^\dagger(\Omega) S_{21}(\Omega), \; S_{12}^\dagger(\Omega) S_{22}(\Omega) \, \in \; \mathbb{R}\; \forall \; \Omega \in \; \mathbb{R} \\
    &S_{11}^\dagger(\Omega) S_{22}(\Omega) - S_{12}^\dagger(\Omega) S_{21}(\Omega) = 1 \; \forall \;\Omega \in \; \mathbb{R}
\end{align}
\label{supp_eqn:symp_trans_conditions}
\end{subequations}
We realize that we can write down a symplectic transformation satisfying these constraints in the following manner,
\begin{align}
&\mathbf{S}(\Omega) = e^{i \phi(\Omega)} \left[\begin{array}{cc}
     A(\Omega) \; & B(\Omega) \\
     C(\Omega) \; & D(\Omega)
\end{array}\right], \nonumber \\ &A(\Omega), B(\Omega), C(\Omega), D(\Omega), \phi(\Omega) \in \, \mathbb{R}\; \forall \; \Omega, \nonumber \\
&A(\Omega)D(\Omega)-B(\Omega)C(\Omega)= 1\;\forall \;\Omega
\end{align}
where the ``phase" $\phi(\Omega)$ ensures causality by canceling out the anti-causal poles of $A(\Omega), B(\Omega), C(\Omega)$ and $D(\Omega)$. Furthermore, we can replace $A(\Omega), B(\Omega), C(\Omega)$ and $D(\Omega)$ with other real functions of $\Omega$ such that
\begin{widetext}
\begin{align}
    \left[\begin{array}{cc}
     A(\Omega) \; & B(\Omega) \\
     C(\Omega) \; & D(\Omega)
\end{array}\right] &= \left[\begin{array}{cc}
\cos{\theta(\Omega)}\cosh{r(\Omega)} + \cos{\rho(\Omega)}\sinh{r(\Omega)} & -\sin{\theta(\Omega)}\cosh{r(\Omega)} + \sin{\rho(\Omega)}\sinh{r(\Omega)}  \\[2mm]
\sin{\theta(\Omega)}\cosh{r(\Omega)} + \sin{\rho(\Omega)}\sinh{r(\Omega)} & \cos{\theta(\Omega)}\cosh{r(\Omega)} - \cos{\rho(\Omega)}\sinh{r(\Omega)}
\end{array}\right]
\end{align}
\end{widetext}
where $r(\Omega), \, \theta(\Omega),\,\rho(\Omega) \in \mathbb{R} \,
\forall \, \Omega$. It is straightforward to check that this form satisfies the constraints in Eq. (\ref{supp_eqn:symp_trans_conditions}).

\section{Realization of frequency-dependent squeezing}
\label{app:freq_dep_squeezing}

We aim to find the parameters of a filter that squeezes the input light field such that the spectrum of the quantum noise in the output is squeezed by the same amount at every frequency, i.e. the shot noise and the quantum radiation pressure noise (QRPN) is squeezed simultaneously \cite{kimble_conversion, lee_ligo_freq_squeezing}. First, we write down the input/output relations for the light field only, 
\begin{align}
\label{supp_eqn:ponderomotive_squeezing}
    \begin{bmatrix}
        \hat v_1(\Omega) \\
        \hat v_2(\Omega)
    \end{bmatrix} = \begin{bmatrix}
        1 & 0 \\[1mm]
        \Omq^2 \, \chi(\Omega) & 1
    \end{bmatrix} \begin{bmatrix}
         \hat u_1(\Omega) \\
        \hat u_2(\Omega)
    \end{bmatrix}
\end{align}
Assuming that the input light field is in the vacuum state, the spectrum of the quantum noise $S(\Omega)$ in the phase quadrature $\hat v_2$ is given by
\begin{align}
    S(\Omega) = 1 + \Omq^4 \, |\chi(\Omega)|^2
\end{align}
We observe that the first term contributes as a shot noise, whereas the second term is the QRPN since it is transduced by the mechanical oscillator. In order to realize frequency-dependent squeezing, we first squeeze the input light field in a frequency-independent manner (for example, with propagation through a bulk crystal), and filter it through a detuned Fabry-Perot cavity. Such an operation will have the following transfer function in the frequency domain
\begin{align} 
   \frac{1}{\widetilde{C}} \begin{bmatrix}
        \Omega^2+\gamma^2-\Delta^2 & -2 \gamma \Delta \\
        2 \gamma \Delta & \Omega^2+\gamma^2-\Delta^2
    \end{bmatrix} \dotproduct \begin{bmatrix}
        e^r & 0 \\
        0 & e^{-r}
    \end{bmatrix}
\end{align}
where $\widetilde{C} = \Delta^2 - (\Omega + i\gamma)^2$. $\gamma$ and $\Delta$ are the cavity linewidth and detuning, respectively, and $r$ is the squeezing factor, with $r, \gamma, \Delta \in \mathbb{R}, \, r, \gamma>0$. Plugging this in Eq. (\ref{supp_eqn:ponderomotive_squeezing}), assuming a high-Q oscillator (i.e. $\omegam \gg \gammam$), $\hat v_2$ will be in the form of
\begin{align}
\label{supp_eqn:v2_after_squeezing}
    \hat v_2 &= \frac{e^r}{\widetilde{C}}\left( -2\gamma\Delta + \frac{\Omq^2 \, (\gamma^2+\Omega^2 - \Delta^2)}{\Omega^2 - \omegam^2} \right) \hat u_1 \nonumber \\
    & +\frac{e^{-r}}{\widetilde{C}}\left( \Delta^2 -\gamma^2-\Omega^2 - \frac{\Omq^2 \, 2 \gamma \Delta}{\Omega^2 - \omegam^2} \right) \hat u_2
\end{align}
where $\hat u_1$, $\hat u_2$ are the vacuum fluctuations. We notice that the coefficient multiplying the amplitude quadrature in Eq. (\ref{supp_eqn:v2_after_squeezing}) can be set to zero with an appropriate choice of $\gamma, \Delta$. More specifically, if we set
\begin{align}
    \gamma = \sqrt{\frac{-\omegam^2 + \sqrt{\omegam^4 + \Omq^4}}{2}}, \quad \Delta = \frac{\Omq^2}{2\gamma},
\end{align}
we obtain that the spectrum of the quantum noise is modified as
\begin{align}
    S(\Omega) \rightarrow e^{-2r} S(\Omega)
\end{align}
Therefore, the spectrum is squeezed with a fixed factor of $e^{-2r}$.

\section{Passive losses}
\label{app:passive_loss}

Here, we consider passive (detection) losses, affecting the readout of the output light field. Such losses occur due to processes such as the finite quantum efficiency of the photodetector. They can be modeled as \cite{Danilishin_qm_theory} 
\begin{align}
    \hat v_i'(t) = \sqrt{\eta} \, \hat v_i(t) + \sqrt{1-\eta} \, \hat e_i(t), \quad i = 1, 2,
\end{align}
where $\hat v'_i(t)$ is the measured quadrature of the output light field, $\hat e_i(t)$ is an additional quantum noise that is assumed to be in the vacuum state, and $0 < \eta < 1$ is the efficiency of the photodetector. Then, $S_{e_i e_i} = 1$, $i = 1, 2$, and $ S_{e_1 e_2} = 0$. We can rewrite the equations of motion in Eqs. (\ref{eqn:eqns_time_domain}) in terms of $\hat v'_{1,2}(t)$  as
\begin{subequations}
\label{eq:passive_losses_1}
\begin{align}
    \hat{v}'_1(t) &= \sqrt{\eta} \, \hat u_1(t) + \sqrt{1-\eta} \, \hat e_1(t), \\
\hat{v}'_2(t) &= \sqrt{\eta} \left[ \hat u_2(t)  + \frac{\Omq}{\sqrt{\omegam}} (\hat{b}_1(t)+\hat{n}_{\rm{S}}(t)) \right] \nonumber \\ &\quad + \sqrt{1-\eta} \, \hat e_2(t) \,, \\
\dot{\hat{b}}_2(t) &= -\gammam \hat{b}_2(t) - \omegam \hat{b}_1(t) + \frac{\Omq}{\sqrt{\omegam}} \hat{u}_1(t) + \hat{n}_{\rm{F}}(t), \\
\dot{\hat{b}}_1(t) &= \omegam \hat{b}_2(t),
\end{align}
\end{subequations}
If the ingoing light field is not frequency-dependently squeezed, such that $\hat u_{1,2}(t)$ denote vacuum fluctuations, we can combine the operators $\hat u_{1,2}(t)$ and $\hat e_{1,2}(t)$ in the form a white noise. In this way, Eqs. (\ref{eq:passive_losses_1}) will resemble the initial equations of motion in the absence of passive losses (Eqs. (\ref{eqn:eqns_time_domain})). Defining $\hat u'_{1, 2}(t) = \sqrt{\eta} \, \hat u_{1,2}(t) + \sqrt{1-\eta} \, \hat e_{1,2}(t)$, we observe that $\hat u'_{1,2}$ is a white noise with $S_{u'_i u'_i} = 1$, $i = 1, 2$, and $ S_{u'_1 u'_2} = 0$, since $\hat u_{1,2}$ and $\hat e_{1,2}$ are independent noise sources. Making sure to preserve the dynamics of the oscillator in terms of the amount of driving noise, we will obtain
\begin{subequations}
\begin{align}
\hat{v}'_1(t) &= \hat u'_1(t), \\
\hat{v}'_2(t) &= \hat u'_2(t)  + \frac{\Omq \sqrt{\eta}}{\sqrt{\omegam}} (\hat{b}_1(t)+\hat{n}_{\rm{S}}(t)) \,, \\
 \dot{\hat{b}}_2(t) &= -\gammam \hat{b}_2(t) - \omegam \hat{b}_1(t) + \frac{\Omq \sqrt{\eta}}{\sqrt{\omegam}} \hat{u}'_1(t) \nonumber \\ & \quad + \hat{n}_{\rm{F}}(t) + \frac{\Omq \sqrt{1 - \eta}}{\sqrt{\omegam}} \hat \xi(t), \\
\dot{\hat{b}}_1(t) &= \omegam \hat{b}_2(t),
\end{align}
\end{subequations}
where $\hat \xi$ is an additional noise in the form of vacuum fluctuations. We notice that $\Omq$ is scaled by $\sqrt{\eta}$, and there appears to be an additional white force noise on the mechanical oscillator with a constant spectrum of $\Omq^2 (1-\eta)/\omegam$, due to the rescaling of $\Omq$ by $\eta$. Note that the total force noise on the oscillator is, as expected, unchanged. Due to the dependency of this additional force noise to the coherent coupling $\Omq$, the entangling-disentangling transition is no longer universal with respect to $\Omq$. 

Then, we encounter two distinct limits: if $S_{\nf}(\Omega) \gg \Omq^2 (1-\eta)/\omegam$, the entangling-disentangling transition is approximately unaffected. However, we expect the amount of entanglement in the system---if there exists any---to reduce because of the rescaling of $\Omq$ by $\eta$. On the other hand, if $S_{\nf}(\Omega) \ll \Omq^2 (1-\eta)/\omegam$, it is more advantageous to have a small coupling $\Omq$ in order to achieve entanglement, as the additional force noise in the system is approximately proportional to $\Omq$.

%\bibliography{refs}


\end{document}